\documentstyle[aps,eqsecnum,floats]{revtex}

\input{psfig.tex}
\begin{document}
\author{Serge Winitzki\thanks{%
swinitzk@cosmos2.phy.tufts.edu}}
\address{Department of Physics and Astronomy, Tufts University,
Medford, Massachusetts~~02155}
\author{Arthur Kosowsky\thanks{%
akosowsky@physics.rutgers.edu}}
\address{Department of Physics and Astronomy, Rutgers University, 136
Frelinghuysen Road,
Piscataway, New Jersey 08854-8019}
\title{Minkowski Functional Description of Microwave Background Gaussianity}
\date{\today }
\maketitle

\begin{abstract}
A Gaussian distribution of cosmic microwave background temperature
fluctuations is a generic prediction of inflation. Upcoming
high-resolution maps of the microwave background will allow detailed
tests of Gaussianity down to small angular scales, providing a crucial
test of inflation. We propose Minkowski functionals as a calculational
tool for testing Gaussianity and characterizing deviations from it. We
review the mathematical formalism of Minkowski functionals of random
fields; for Gaussian fields the functionals can be calculated exactly.
We then apply the results to pixelized maps, giving explicit expressions
for calculating the functionals from maps as well as the Gaussian
predictions, including corrections for map boundaries, pixel noise, and
pixel size and shape. Variances of the functionals for Gaussian
distributions are derived in terms of the map correlation
function. Applications to microwave background maps are discussed.
\end{abstract}


\section{Introduction}

The Cosmic Microwave Background (CMB) provides the earliest obtainable
direct information about the Universe. Upcoming experiments (see, e.g., MAP
1996; Planck 1996) will map CMB temperature and polarization fluctuations
with unprecedented sensitivity and angular resolution. Recent analyses have
demonstrated that such maps contain a wealth of cosmological information,
allowing a high-precision determination of the fundamental cosmological
parameters (Jungman et al. 1996a,b; Zaldarriaga et al. 1997; Bond et al.
1997; Kamionkowski and Kosowsky 1997) and a clean separation of inflationary
and topological defect models of structure formation (Pen et al. 1997; Hu
and White 1997). These exciting prospects will keep the CMB at the forefront
of cosmology for the coming decade.

Most theoretical and experimental results to date have focussed on the
angular power spectrum of the CMB. If the temperature and polarization
fluctuations are Gaussian, then the power spectrum contains all available
information about the fluctuations; even if the fluctuations depart from
Gaussian, the power spectrum is still a valuable and fundamental
characterization of the fluctuations. For experiments on large angular
scales, notably the DMR maps from the COBE satellite (Bennett et al. 1996),
homogeneity of the Universe essentially guarantees a Gaussian anisotropy
distribution via the central limit theorem since the experiment only probes
scales outside of the causal horizon at last scattering (for a relevant
discussion, see Scherrer and Schaefer 1995). Probes of the anisotropies at
resolutions of one degree and smaller hold the possibility of uncovering
non-Gaussian signatures. Theoretically, Gaussian fluctuations are well
motivated, as inflation generically produces Gaussian-distributed scalar and
tensor metric perturbations (Guth and Pi 1982; Starobinskii 1982; Hawking
1982; Bardeen et al. 1983), although this prediction can be circumvented in
some unusual models (Allen et al. 1987; Salopek et al. 1989). Measurement of
Gaussian fluctuation statistics at all angular scales would clear a major
hurdle for inflation; conversely, discovery of cosmological non-Gaussianity
would cast strong doubt on the inflationary scenario. If the Universe is not
described by an inflationary model, non-Gaussianity will be an important
clue in unraveling the nature of the primordial density perturbations.

A variety of techniques have been applied so far to test for Gaussianity in
CMB data, primarily COBE maps. The two-point correlation function is
equivalent to the fluctuation power spectrum. In principle, the higher-order
correlation functions provide a complete test of Gaussianity; indeed, a
Gaussian distribution can be defined as one in which the odd (3-point,
5-point, etc.) correlations are all zero, while the even correlation
functions all factorize into products of two-point functions. Kogut et al.
(1996), following earlier analysis of the 2-year maps (Smoot et al. 1994)
and theoretical work (Luo and Schramm 1993; Luo 1994), have calculated
certain 3-point functions for the COBE 4-year maps, finding the data
consistent with zero (Gaussian). A drawback of analyzing N-point correlation
functions is that computation time becomes prohibitive for higher-point
functions unless restricted to specific fixed configurations of points,
which provides a weaker test of Gaussianity than the most general case.
Also, higher-point analysis of maps with complicated boundaries presents
further difficulties.

Another, more heuristic statistic is the two-point correlation of
temperature extrema (peaks and valleys), introduced by Bond and Efstathiou
(1987), who provide analytic approximations to the correlation function for
noiseless Gaussian random fields. The COBE 4-year maps are also consistent
with Gaussianity for this statistic (Kogut et al. 1996). Finally,
substantial attention has been focussed on the topological genus of
temperature contours (Coles, 1988; Gott et al. 1990; Smoot et al. 1994;
Torres et al. 1995; Colley et al. 1996; Kogut et al. 1996). The genus, as
will be discussed below, is equivalent to one of the three Minkowski
functionals for a two-dimensional map.

In this paper we concentrate on the set of statistics known as Minkowski
functionals (Minkowski 1903). A general theorem of integral geometry states
that all properties of a $d$-dimensional convex set (or more generally, a
finite union of convex sets) which satisfy translational invariance and
additivity
(called {\em morphological} properties)
are contained in $d+1$ numerical values (Hadwiger, 1956 and
1959). For a pixelized temperature map $T({\bf n})$ (or, e.g., a density
field), we consider the excursion sets of the map, defined as the set of all
map pixels with value of $T$ greater than some threshold $\nu $ (see, e.g.,
Weinberg et al. 1987; Melott 1990). Then the three functionals of these
excursion sets completely describe the morphological properties of the
underlying temperature map $T({\bf n})$. In two and three dimensions, the
Minkowski functionals have intuitive geometric interpretations. The key
point relevant for this paper is that the Minkowski functionals can be
calculated exactly for an underlying Gaussian field (Tomita 1990). Thus they
provide a natural set of statistical tests for Gaussianity. Recently,
Buchert and collaborators have introduced Minkowski functionals into the
cosmological literature, primarily in the context of three-dimensional point
distributions applied to galaxy surveys (Mecke et al. 1994; Buchert 1995;
Kerscher et al. 1997).

For a two-dimensional map, the three Minkowski functionals correspond
geometrically to the total fractional area of the excursion set, the
boundary length of the excursion set per unit area, and the Euler
characteristic per unit area (equivalent to the topological genus). The area
(Coles 1988, Ryden et al. 1989) 
and the genus (mentioned above) have been previously considered
in the context of microwave background temperature maps, but not in the
unified formal context of the Minkowski functionals. The additivity
property, previously not exploited, allows an evaluation of the standard
errors in estimating the functionals from a map, rather than resorting to
Monte Carlo simulations as in previous work (Scaramella and Vittorio 1991;
Torres et al. 1995; Kogut et al. 1996). A second property, given by the
principle kinematic formulas, provides a recipe for incorporating arbitrary
map boundaries in a straightforward way, a valuable advantage when analyzing
upcoming maps with irregular portions of the sky cut out to eliminate
contamination from foreground emission.

This paper applies the formalism of Minkowski functionals to pixelized
two-dimensional maps as a test of Gaussianity; the results will be directly
applicable to temperature maps of the microwave background. In Sec.~II,
we present an overview of Minkowski functionals of continuous fields and
explicate their relevant mathematical properties. Section III then considers
the related case of Minkowski functionals on a lattice, or equivalently of a
pixelized map. We give explicit formulas for variances of the functionals
and corrections for boundaries. We also give explicit calculational
algorithms for different pixelization schemes. Section IV addresses the
issue of Gaussianity, giving the explicit forms for the Minkowski
functionals for underlying Gaussian statistical distributions. Pixelization
introduces corrections to the Gaussian functionals in the smooth case, which
we derive as power series in the pixel size. Finally, we conclude in section
V with a discussion of various issues related to analyzing maps, including
noise, smoothness, and numerical efficiency.

\section{Review of Minkowski Functionals}

We are interested in characterizing the morphological properties of
microwave background maps, the properties of the map which are invariant
under translations and rotations and which are additive. For example, the
area of the map which is above a certain temperature threshold is a
morphological characteristic. The branch of mathematics known as integral
geometry provides a natural tool for this characterization, known as the
Minkowski functionals (or {\em quermassintegrals}). In this section we give
a brief review of Minkowski functionals and summarize their basic
properties. For further details and general mathematical background, see
Weil (1983) or Stoyan et al. (1987).

Consider a convex set $K$ in ${\bf R}^d$. The parallel set of distance $r$
to $K$ is the set 
\begin{equation}
K_r=\bigcup_{k\in K}b(k,r),  \label{parallelset}
\end{equation}
where $b(k,r)$ is the closed ball of radius $r$ centered at the point $k$. A
relation called the Steiner formula can be taken as the definition of the
Minkowski functionals $W_i$: 
\begin{equation}
V(K_r)=\sum_{i=0}^d%
{d \choose i}
W_i(K)r^i,  \label{steiner}
\end{equation}
where $V$ denotes the ($n$-dimensional) volume. For low-dimensional spaces,
the Minkowski functionals can be expressed simply in terms of the
geometrical quantities length $\ell (K)$ (for $d=1$), area $A(K)$ and
boundary length $L(K)$ (for $d=2$), and volume $V(K)$, surface area $S(K)$,
and mean breadth ${\bar{b}}(K)$ (for $d=3$): 
\begin{mathletters}
\begin{equation}
W_0(K)=\ell (K),\qquad W_1(K)=2\qquad (d=1);  \label{onedim}
\end{equation}
\begin{equation}
W_0(K)=A(K),\qquad W_1(K)={\frac 12}L(K),\qquad W_2(K)=\pi \qquad (d=2);
\label{twodim}
\end{equation}
\begin{equation}
W_0(K)=V(K),\qquad W_1(K)={\frac 13}S(K),\qquad W_2(K)={\frac{2\pi }3}{\bar{b%
}}(K),\qquad W_3(K)={\frac{4\pi }3}\qquad (d=3).  \label{threedim}
\end{equation}
The Steiner formula can also be generalized to the other Minkowski
functionals besides volume as 
\end{mathletters}
\begin{equation}
W_j(K_r)=\sum_{i=0}^{d-j}%
{d-j \choose i}
W_{i+j}(K)r^i.  \label{steinergen}
\end{equation}

The motivation for using the Minkowski functionals for characterizing
morphology is the following completeness theorem of Hadwiger (1959). 
For ${\cal K}%
^d $ the class of convex, compact sets in ${\bf R}^d$, consider a continuous
map $T:{\cal K}^d\rightarrow {\bf R}$ which satisfies the properties of
motion invariance and additivity: 
\begin{equation}
T(gK)=T(K)\,\,\,\forall K\in {\cal K}^d,\,\,g\in G_d,  \label{invariance}
\end{equation}
where $G_d$ is the group of rigid motions in $d$ dimensions (i.e., rotations
and translations); and 
\begin{equation}
T(K_1\cup K_2)+T(K_1\cap K_2)=T(K_1)+T(K_2)  \label{additivity}
\end{equation}
for $K_1,\,K_2\in {\cal K}^n$ with $K_1\cup K_2\in {\cal K}^d$. Then $T$ can
be expressed as a linear combination of the Minkowski functionals: 
\begin{equation}
T(K)=\sum_{i=0}^d\alpha _iW_i(K),\quad \alpha _i\in {\bf R}.
\label{hadwiger}
\end{equation}
In other words, all of the morphological information about a convex body is
contained in the Minkowski functionals.

In the following sections, we will not be concerned with characterizing a
single convex set but rather a finite union of convex sets (i.e.\ map
pixels), or in mathematical terms, sets which are elements of the convex
ring ${\cal R}^n$ (Hadwiger 1956 and 1959). In two dimensions, the area and
boundary length functionals have obvious generalizations to the convex ring,
while the third functional becomes the Euler-Poincare characteristic. The
Hadwiger characterization theorem, Eq.~(\ref{hadwiger}), generalizes to the
convex ring. In this paper, we will find it convenient to work with a
different normalization of the functionals, given by 
\begin{equation}
M_i(K)\equiv \frac{\omega _{d-i}}{\omega _d\omega _i}W_i(K),  \label{Mdef}
\end{equation}
where $\omega _d=\pi ^{d/2}/\Gamma (1+d/2)$ is the volume of the unit ball
in $d$ dimensions. The three functionals of an element of ${\cal R}^2$ are
then 
\begin{equation}
M_0 =A, \qquad M_1 =\frac 1{2\pi }L,\qquad M_2 =\frac 1\pi \chi ,
\label{Mexpl}
\end{equation}
where $\chi$ is the Euler characteristic.

Another set of useful relations are the so-called principle kinematic formulae
(Hadwiger 1957, Santalo 1976), which can be written concisely as (Mecke et
al. 1994) 
\begin{equation}
\int_{G_d}M_i(K_1\cap gK_2)dg=\sum_{j=0}^i%
{i \choose j}
M_{i-j}(K_2)M_j(K_1).  \label{kinematic}
\end{equation}
The integral is over the group of motions, formally with an invariant Haar
measure $dg$. This formula describes the set $K_1$ through its intersections
with a test body $K_2$ in random orientations.

To analyze a map in terms of Minkowski functionals, we consider the
excursion sets of the map, the map subset which exceeds a fixed threshold
value. The threshold is treated as an independent variable on which the
Minkowski functionals of the excursion set depend. The three functionals of
interest are (up to irrelevant constant factors, cf.\ Eq.~(\ref{twodim}))
the area of the excursion set $A$, its boundary length $L$, and its
Euler-Poincare characteristic $\chi$ (or equivalently, its topological
genus). The following section discusses in detail how to construct a
discrete version of these functionals, applicable to a pixelized map.

\section{Minkowski functionals on a lattice}

\label{Sec:MinkLattice}

The method of analysis proposed in this paper consists of comparing the
values of the Minkowski functionals $M_i $ calculated from an experimentally
obtained map with the theoretical predictions for the expectation values and
variances of $M_i $ on Gaussian distributions. Carrying out this program
requires calculation of the Minkowski functionals from given maps, properly
taking into account the boundary of the observed region and the effect of
pixelization.

This section presents the necessary formalism for the analysis of Minkowski
functionals on a lattice. Rather than treat pixelized maps as approximations
to the ``true'' continuous temperature field, we apply the formalism of the
Minkowski functionals directly to random functions on discrete lattices
(for extensive mathematical background, see Serra 1982).
Although some elements of this consideration are present in the literature
(Hamilton et al. 1986, Coles 1988, Likos et al. 1995), we give an
independent and self-contained derivation of our results for two-dimensional
Euclidean maps. The general expressions for the expectation values and
variances of the Minkowski functionals are calculated for a homogeneous and
isotropic random function on a regular lattice, in terms of the probability
distributions for the random function. This will be the foundation for the
analysis of Sec.~\ref{Sec:Gaussian}. Explicit calculational algorithms
for the area, boundary length, and Euler characteristic are given. We also
derive formulas for the boundary corrections to these Minkowski functionals.

\subsection{General formalism}

Consider a homogeneous and isotropic scalar function $f$ given on some
regular lattice in a region $S$ of a $2$-dimensional plane, so that each
lattice element (pixel) $s$ is assigned a number $f\left( s\right) $. We
define the excursion set $S_u$ of the function $f$ at level $u$ as the union
of all pixels $s\in S$ for which $f\left( s\right) >u$. For instance, the
lattice may be a regular square lattice with a given step $\Delta $, and the
pixels $s$ would then be squares with side $\Delta $; the region $S_u$ would
then consist of all squares $s$ for which $f\left( s\right) >u$. Another
possibility is a hexagonal lattice made up of regular hexagons. Defining the
indicator function $N_u\left( s\right) $ as $N_u\left( s\right) =1$ if $%
f\left( s\right) >u$ and $N_u\left( s\right) =0$ otherwise, the excursion
set is symbolically represented as a union
\begin{equation}
S_u=\bigcup_{s\in S}sN_u\left( s\right) ,  \label{AUnion}
\end{equation}
where implicitly only terms with $N_u\left( s\right) =1$ are present in the
union. Note that a temperature map may conveniently be converted to
dimensionless units by expressing the temperature deviation from the mean in
units of the root-mean-square temperature deviation of the map.

To characterize a given map $f$, we use the values of Minkowski functionals $%
M_i$ on the excursion set $S_u$ generated from $f$ at a given level $u$.
Because of additivity of Minkowski functionals, Eq.~(\ref{additivity}), the
following decomposition formula follows from Eq.\ (\ref{AUnion}): 
\begin{eqnarray}
M\left( S_u\right) &=&\sum_{s\in S}N_u\left( s\right) M\left( s\right)
-\sum_{s_1\neq s_2}N_u\left( s_1\right) N_u\left( s_2\right) M\left( s_1\cap
s_2\right)  \nonumber \\
&&\ +\sum_{s_1\neq s_2\neq s_3}N_u\left( s_1\right) N_u\left( s_2\right)
N_u\left( s_3\right) M\left( s_1\cap s_2\cap s_3\right) -\ldots ,
\label{MSum}
\end{eqnarray}
where the sums are taken over all different pairs, triples etc.\ of lattice
elements $s$. The intersections $s_1\cap s_2\cap ...$ of pixels are
understood in the simple geometric sense, as intersections of polygons. Note
that it is only necessary to sum over adjacent pixels in Eq.\ (\ref{MSum}),
since the Minkowski functionals $M$ vanish on empty sets. Therefore, the
series in Eq.\ (\ref{MSum}) is actually finite and stops when the
intersection of the lattice elements is empty and $M\left( s_1\cap s_2\cap
...\cap s_n\right) =0$. For instance, the intersection of any two or more
pixels has zero area, so the series (\ref{MSum}) for the area functional $%
M_0\equiv A$ stops after the first term. In a square lattice, at most four
squares can have non-empty intersection; in a hexagonal lattice, at most
three hexagons intersect, and the series (\ref{MSum}) stops after fewer
terms. For this reason, it is generally simpler to analyze the Minkowski
functionals on lattices with hexagonal symmetry.

Equation (\ref{MSum}) expresses the variable $M\left( S_u\right) $ through
the function $N_u\left( s\right) $ and the constants $M\left( s\right) $, $%
M\left( s_1\cap s_2\right) $, ..., which are trivially calculated for any
given lattice geometry (these are just the Minkowski functionals applied to
individual pixels and their intersections, determined by the area and side
lengths of the pixels). Direct application of Eq.\ (\ref{MSum}) can be used
to calculate the Minkowski functionals $M\left( S_u\right) $ for a given map
at a given level $u$ by simply evaluating $N_u$ $\left( s\right) $ for each
pixel $s$. Note that despite the sums over all pairs, triples and so on
appearing in Eq.\ (\ref{MSum}), in fact only terms with adjacent pixels give
nonzero contributions, and the required computation time is linear in the
total number of pixels. This will also be clear from the explicit
computational algorithms below.

We now consider expectation values and variances of $M\left( S_u\right) $.
Taking expectation values of both sides of Eq.\ (\ref{MSum}), we obtain 
\begin{equation}
\left\langle M\left( S_u\right) \right\rangle =\sum_{s\in S}\left\langle
N_u\left( s\right) \right\rangle M\left( s\right) -\sum_{s_1\neq
s_2}\left\langle N_u\left( s_1\right) N_u\left( s_2\right) \right\rangle
M\left( s_1\cap s_2\right) +\ldots  \label{AveMSum}
\end{equation}
The averages $\left\langle N_u\left( s\right) \right\rangle $, $\left\langle
N_u\left( s_1\right) N_u\left( s_2\right) \right\rangle $, ... in Eq.\ (\ref
{AveMSum}) must be calculated for a given random function $f$. Homogeneity
and isotropy of $f$ leads to a simplification of Eq.\ (\ref{AveMSum}); for
instance $\left\langle N_u\left( s\right) \right\rangle $ will not depend on 
$s$, and $\left\langle N_u\left( s_1\right) N_u\left( s_2\right)
\right\rangle $ will be a function of the distance between $s_1$ and $s_2$
only.

Similarly, the variances of the functionals $M\left( S_u\right) $ can be
found by substituting the decomposition formula (\ref{MSum}) into the
general expression 
\begin{equation}
\text{var}\left[ M\right] \equiv \left\langle M\left( S_u\right)
^2\right\rangle -\left\langle M\left( S_u\right) \right\rangle ^2.
\label{VarM}
\end{equation}
For instance, the variance of the area $A$ is 
\begin{equation}
\text{var}\left[ A\right] =A_1^2\left( \sum_{s_1,s_2}\left\langle N_u\left(
s_1\right) N_u\left( s_2\right) \right\rangle -\left[ \sum_s\left\langle
N_u\left( s\right) \right\rangle \right] ^2\right) ,  \label{VarArea}
\end{equation}
where $A_1$ is the area of one pixel. The variance will in general come from
two contributions, the ``cosmic variance'' due to sampling of an underlying
random field, and the noise variance arising from pixel noise. We shall
incorporate the pixel noise into the distribution for the random field (see
Sec.~\ref{Sec:Gaussian}) and treat the variance as exclusively a result
of sampling of the underlying ensemble. The general expression (\ref{VarM})
can in principle be used to compute the variances of the Minkowski
functionals analytically for given field distributions. We give a
calculation for the variance of the area for a Gaussian distributed map in
Sec.~\ref{Sec:Gaussian}, along with approximations for the
variances of the boundary length and Euler characteristic.

A homogeneous random field may be defined by joint $n$-point distribution
densities $p_n\left( x_1,...,x_n;s_1,...,s_n\right) $ for its values $x_i$
on some given configurations of $n$ points $s_1$, ..., $s_i$. Due to
homogeneity, the densities $p_n$ depend only on distances between the points 
$s_i$, and we write them as $p_1\left( x\right) $, $p_2\left(
x_1,x_2;r_{12}\right) $, $p_3\left( x_1,x_2,x_3;r_{12},r_{23},r_{31}\right) $
and so on. The averages of products of the indicator functions $N_u\left(
s\right) $, which will frequently arise in our calculations, can be
expressed through integrals of $p_n$. For instance, the average $%
\left\langle N_u\left( s\right) \right\rangle $ is related to the one-point
distribution density $p_1\left( x\right) $ as 
\begin{equation}
\left\langle N_u\left( s\right) \right\rangle =\int_u^\infty p_1\left(
x\right) dx\equiv P_1\left( u\right) .  \label{P1Def}
\end{equation}
Here we have introduced the cumulative distribution function $P_1\left(
u\right) $; it does not depend on $s$ due to homogeneity of the field.
Similarly, the average $\left\langle N_u\left( s_1\right) N_u\left(
s_2\right) \right\rangle $ for a pair $\left( s_1,s_2\right) $ of points is
equal to the probability that $f\left( s\right) >u$ at both points and
depends only on the distance $r$ between them: 
\begin{equation}
\left\langle N_u\left( s_1\right) N_u\left( s_2\right) \right\rangle
=\int_u^\infty dx_1\int_u^\infty dx_2\ p_2\left( x_1,x_2;r\right) \equiv
P_2\left( u;r\right) .  \label{P2Def}
\end{equation}
In the same manner, one can express the average of a product of the
indicator functions at any $n$ points $\left\langle N_u\left( s_1\right)
...N_u\left( s_n\right) \right\rangle $ through the corresponding $n$-point
distribution function $P_n\left( u;\left| s_i-s_j\right| \right) $, 
\begin{equation}
\left\langle N_u\left( s_1\right) ...N_u\left( s_n\right) \right\rangle
=\int_u^\infty dx_1...\int_u^\infty dx_n\ p_2\left(
x_1,...,x_n;r_{ij}\right) \equiv P_n\left( u;r_{ij}\right) .  \label{PNDef}
\end{equation}
Note that derivation of Eq.\ (\ref{P2Def}) depends on the assumption that
every pixel has the same number of neighbors and so disregards the boundary
pixels. The same limitation applies to Eq.\ (\ref{PNDef}). The effect of the
boundary will be considered separately below.
Also notice that that the distribution functions $P_n$ depend directly
on the $n$-point distribution densities $p_n$ of the field and therefore in
general cannot be reduced to the moments of the field. The important
exception is the case of a Gaussian random field which is considered in
Sec.~\ref{Sec:Gaussian}.

The following subsections consider each of the three Minkowski functionals
in turn: the area, the boundary length, and the Euler characteristic. The
value of each functional will be normalized to the total area of the map $%
A(S)$.

\subsection{The area}

The area functional $M_0\equiv A$ is the simplest of the three. To calculate
the value of $A$ for a given map with $N$ pixels, we count all pixels $s$
with values $f\left( s\right) >u$ (i.e.\ with $N_u\left( s\right) =1$) and
divide by the total number of pixels: 
\begin{equation}
\frac{A\left( S_u\right) }{A\left( S\right) }=\frac 1N\sum_sN_u\left(
s\right) .  \label{AreaN}
\end{equation}
This agrees with Eq.\ (\ref{MSum}) since, as noted above, only the first
term survives in that series. The expectation value of the area functional
is then easily found: 
\begin{equation}
\frac{\left\langle A\left( S_u\right) \right\rangle }{A\left( S\right) }%
=P_1\left( u\right) .  \label{NormArea}
\end{equation}

The variance of the area functional is given by Eq.\ (\ref{VarArea}), which
can be re-written as 
\begin{equation}
\text{var}\left[ \frac{A\left( S_u\right) }{A\left( S\right) }\right] =\frac 
1{N^2}\sum_{s_1,s_2}\left\langle N_u\left( s_1\right) N_u\left( s_2\right)
\right\rangle -\left[ P_1\left( u\right) \right] ^2.  \label{VarArea1}
\end{equation}
In terms of the distribution $P_2\left( u;r\right) $ defined by Eq.\ (\ref
{P2Def}), the double sum in Eq.\ (\ref{VarArea1}) becomes 
\begin{equation}
\frac 1{N^2}\sum_{s_1,s_2}\left\langle N_u\left( s_1\right) N_u\left(
s_2\right) \right\rangle =\frac 1{N^2}\sum_{s_1,s_2}P_2\left( u;\left|
s_1-s_2\right| \right) .
\end{equation}
For small lattice steps, the last sum can be approximated by an integral 
\begin{equation}
\frac 1{N^2}\sum_{s_1,s_2}P_2\left( u;\left| s_1-s_2\right| \right) \approx 
\frac 1{N\Delta ^2}\int_0^{r_{\max }}P_2\left( u;r\right) 2\pi rdr,
\label{P2SumInt}
\end{equation}
where the integral replaces the summation over steps of $\Delta $, and $%
r_{\max }$ is the maximum distance between points of the observed region. We
assumed in Eq.\ (\ref{P2SumInt}) that every pixel $s_1$ is surrounded by a
circular region of size $r_{\max }$ and area $\pi r_{\max }^2=N\Delta ^2$ in
which the pixels $s_2$ is chosen, thereby disregarding the effect of the
actual boundary of the region $S$. The existence of the boundary
affects the
integral in Eq.\ (\ref{P2SumInt}) only at large $r$, and it will be shown
momentarily that the variance does not depend on the behavior at the upper
limit $r_{\max }$, as long as the correlations between points become
negligible at such distances.

The variance of the area functional becomes 
\begin{equation}
\text{var}\left[ \frac{A\left( S_u\right) }{A\left( S\right) }\right] =\frac 
1{\pi r_{\max }^2}\int_0^{r_{\max }}P_2\left( u;r\right) 2\pi rdr-\left[
P_1\left( u\right) \right] ^2.  \label{VarArea2}
\end{equation}
The analytic form of the distribution function $P_2\left( u;r\right) $ is
usually difficult to obtain because the integrals in Eq.\ (\ref{P2Def})
cannot be evaluated. However, it is possible to draw general conclusions
about the variance of $A$ from Eq.\ (\ref{VarArea2}). Since correlations
between distant points are absent for physical reasons, the probability $%
P_2\left( u;r\right) $ of two points to simultaneously have values of $f$
greater than $u$ is approximately equal to $\left[ P_1\left( u\right)
\right] ^2$ at large $r$. If we denote by $r_0$ the distance at which
correlations between points become insignificant so that $P_2\left( u;r\right)
\approx \left[ P_1\left( u\right) \right] ^2$, then we can split the
integration above to obtain 
\begin{eqnarray}
\text{var}\left[ \frac{A\left( S_u\right) }{A\left( S\right) }\right] &=&%
\frac 1{\pi r_{\max }^2}\left( \int_0^{r_0}+\int_{r_0}^{r_{\max }}\right)
2\pi rdr\left\{ P_2\left( u;r\right) -\left[ P_1\left( u\right) \right]
^2\right\}  \nonumber \\
&\approx &\frac 1{A\left( S\right) }\int_0^{r_0}2\pi rdr\left\{ P_2\left(
u;r\right) -\left[ P_1\left( u\right) \right] ^2\right\} .  \label{VarArea3}
\end{eqnarray}
One can obtain an upper bound on the variance (\ref{VarArea3}) by noting
that $P_2\left( u;r\right) \leq P_1\left( u\right) $, which gives 
\begin{equation}
\text{var}\left[ \frac{A\left( S_u\right) }{A\left( S\right) }\right] \leq 
\frac{\pi r_0^2}{A\left( S\right) }P_1\left( u\right) \left[ 1-P_1\left(
u\right) \right] .  \label{VarAreaQ}
\end{equation}
Although this upper bound can only serve as a rough order-of-magnitude
estimate of the variance, it can be used to visualize the overall behavior
of the variance. The peak magnitude of the variance (at $u=0$) is of the
order $r_0^2/A\left( S\right) $ and is small if the size of the observed
region $S$ is much larger than the distance $r_0$ at which correlations
become negligible. The same general conclusion is valid for the variance of
the two other Minkowski functionals. For the case of Gaussian random fields,
we can make further progress in evaluating the variances (see Sec.~\ref
{Sec:Gaussian}).

\subsection{The boundary length}

Now we calculate the boundary length functional $L$ of the set $S_u$,
normalized to the area $A\left( S\right) $ of the observed region $S$. We
shall use Eq.\ (\ref{MSum}), in which we only need to calculate the first
two sums since the intersection of more than two pixels has zero length. The
only intersection of two pixels that has nonzero length is the intersection
of two immediately adjacent pixels that have a common side. So the second
sum in (\ref{MSum}), 
\begin{equation}
\sum_{s_1\neq s_2}N_u\left( s_1\right) N_u\left( s_2\right) L\left( s_1\cap
s_2\right) ,
\end{equation}
needs to be taken only over pairs $\left( s_1,s_2\right) $ of adjacent
pixels with a common side. The number of such immediate neighbors of a given
pixel and the lengths of the corresponding sides depends on the lattice
geometry. If we assume that the pixels are regular polygons with $n_s$ sides
of equal length $L_s$, then $L\left( s_1\cap s_2\right) =2L_s$ (recall that
the boundary length functional of a line segment is equal to {\em twice} the
length of the segment) and the sum (\ref{MSum}) becomes 
\begin{equation}
L\left( S_u\right) =n_sL_s\sum_{s\in S}N_u\left( s\right) -2L_s\sum_{s_1\cap
s_2\neq \emptyset }N_u\left( s_1\right) N_u\left( s_2\right) .  \label{LSum0}
\end{equation}
Normalizing to the total area of the region $A\left( S\right) =NA_1$ gives 
\begin{equation}
\frac{L\left( S_u\right) }{A\left( S\right) }=\frac{n_sL_s}{NA_1}\sum_{s\in
S}N_u\left( s\right) -\frac{L_s}{NA_1}\sum_{s\in S}\sum_{s^{\prime }\text{
adj. to }s}N_u\left( s\right) N_u\left( s^{\prime }\right) .  \label{LSum}
\end{equation}
where the second summation is performed over $n_s$ immediately adjacent
neighbors $s^{\prime }$ of $s$, and we divide by $2$ because of overcounting
of pairs. The formula (\ref{LSum}) can be directly used to calculate the
boundary length functional of a given map at a given level $u$. The required
computation time is again linear in the total number of pixels.

Now consider the expectation value of $L$. We shall use Eq.\ (\ref{LSum})
and the two-point distribution function $P_2\left( u;r\right) $ introduced
in Eq.\ (\ref{P2Def}). If we denote the distance between the centers of any
two adjacent pixels by $\Delta $, then Eq.\ (\ref{LSum}) will give 
\begin{eqnarray}
\frac{\left\langle L\left( S_u\right) \right\rangle }{A\left( S\right) } &=&%
\frac{n_sL_s}{A_1}\frac 1N\sum_{s\in S}\left( \left\langle N_u\left(
s\right) \right\rangle -\frac 1{n_s}\sum_{s^{\prime }\text{ adj. to }%
s}\left\langle N_u\left( s\right) N_u(s^{\prime })\right\rangle \right) 
\nonumber \\
&=&\frac{n_sL_s}{A_1}\left[ P_1\left( u\right) -P_2\left( u;\Delta \right)
\right] .  \label{LAns1}
\end{eqnarray}
Although this expression seems to depend on the lattice geometry, the ratio $%
n_sL_s/A_1$ is the same for any regular lattice which contains only $n_s$%
-sided regular polygons whose centers are separated by the distance $\Delta $%
. For instance, the square lattice is characterized by $n_s=4$, $L_s=\Delta $
and $A_1=\Delta ^2$, which gives $n_sL_s/A_1=4/\Delta $; the hexagonal
lattice with $n_s=6$, $L_s=2\Delta /\sqrt{3}$ and $A_1=\Delta ^2\sqrt{3}$
gives the same answer. This allows us to write the result for all regular
lattices as 
\begin{equation}
\frac{\left\langle L\left( S_u\right) \right\rangle }{A\left( S\right) }=%
\frac 4\Delta \left[ P_1\left( u\right) -P_2\left( u;\Delta \right) \right] .
\label{LAns}
\end{equation}

Since Eq.\ (\ref{LAns}) exhibits an explicit dependence on the lattice step $%
\Delta $, it is important to understand its scaling properties at small $%
\Delta $. It is clear that, for a continuous random field $f$, the
distributions at nearby points are highly correlated, and 
\begin{equation}
\lim_{\Delta \rightarrow 0}P_2\left( u;\Delta \right) =P_1\left( u\right) ,
\end{equation}
which means that both the numerator and the denominator in Eq.\ (\ref{LAns})
will tend to zero at small $\Delta $. However, the exact value of that limit
depends on the behavior of the two-point distribution function $P_2\left(
u;\Delta \right) $ of the random field at small distances $\Delta $. As will
be shown in Sec.~\ref{Sec:Gaussian}, the limit of Eq.\ (\ref{LAns}) as $%
\Delta \rightarrow 0$ for a Gaussian random field is finite if the random
field is ``smooth'' as defined below.

The general expression for the variance with help of Eq.\ (\ref{LSum}) gives
\begin{eqnarray}
\text{var}\left[ \frac{L\left( S_u\right) }{A\left( S\right) }\right]
&=&\left\langle \left( \frac{L\left( S_u\right) }{A\left( S\right) }\right)
^2\right\rangle -\left\langle \frac{L\left( S_u\right) }{A\left( S\right) }%
\right\rangle ^2  \nonumber \\
&=&\left( \frac{n_sL_s}{A_1N}\right) ^2\left\langle \left[ \sum_{s\in
S}N_u\left( s\right) -\sum_{s\in S}\frac 1{n_s}\sum_{s^{\prime }\text{ adj.
to }s}N_u\left( s\right) N_u\left( s^{\prime }\right) \right] ^2\right\rangle
\nonumber \\
&&\qquad\qquad
-\left( \frac{n_sL_s}{A_1N}\right) ^2\left[ \left\langle \sum_{s\in
S}N_u\left( s\right) -\sum_{s\in S}\frac 1{n_s}\sum_{s^{\prime }\text{ adj.
to }s}N_u\left( s\right) N_u\left( s^{\prime }\right) \right\rangle \right]
^2.  \label{VarLSum}
\end{eqnarray}
This expression contains terms such as $\left\langle N_u\left( s_1\right)
N_u\left( s_2\right) N_u\left( s_2^{\prime }\right) \right\rangle $, which
are summed over all pixels $s_1$ and $s_2$ as well as over all nearest
neighbors $s_2^{\prime }$ of $s_2$. Although one can express such averages
through the corresponding distribution $P_3\left( u;r_{ij}\right) $, the
distances $r_{ij}$ between points depend on their relative orientation in a
complicated way, which makes an exact evaluation of Eq.\ (\ref{VarLSum})
difficult. However, an approximate expression for such terms
can be obtained by noting
that for most $3$-point configurations in question the distance
between $s_1$ and $s_2$ is much greater than the distance $\Delta $ between
the nearest neighbor pair $\left( s_2,s_2^{\prime }\right) $. This suggests
that we should disregard the difference between $\left| s_1-s_2\right| $ and 
$\left| s_1-s_2^{\prime }\right| $ and in the first approximation treat all $%
3$-point configurations as equilateral triangles with sides $\left|
s_1-s_2\right| $, $\left| s_1-s_2\right| $ and $\Delta $. Denoting
the $3$-point distribution $P_3$ for such triangles by $P_3\left( u;\Delta
,\left| s_1-s_2\right| \right) $, we have approximately
\begin{equation}
\left\langle N_u\left( s_1\right) N_u\left( s_2\right) N_u\left( s_2^{\prime
}\right) \right\rangle \approx P_3\left( u;\Delta ,\left| s_1-s_2\right|
\right) .  \label{P3Approx}
\end{equation}
Similarly, the terms with products of four $N_u\left( s\right) $ can be
approximated by symmetric $4$-point distribution functions 
\begin{equation}
\left\langle N_u\left( s_1\right) N_u\left( s_2\right) N_u\left( s_1^{\prime
}\right) N_u\left( s_2^{\prime }\right) \right\rangle \approx P_4\left(
u;\Delta ,\left| s_1-s_2\right| \right) ,  \label{P4Approx}
\end{equation}
and then Eq.\ (\ref{VarLSum}) can be rewritten as 
\begin{eqnarray}
\text{var}\left[ \frac{L\left( S_u\right) }{A\left( S\right) }\right]
&\approx &\left( \frac{n_sL_s}{A_1N}\right) ^2\sum_{s_1,s_2}\left[ P_2\left(
u;r_{12}\right) -2P_3\left( u;\Delta ,r_{12}\right) +P_4\left( u;\Delta
,r_{12}\right) \right]  \nonumber \\
&&-\left( \frac{n_sL_s}{A_1}\right) ^2\left[ P_1\left( u\right) -P_2\left(
u;\Delta \right) \right] ^2,
\end{eqnarray}
where $r_{12}\equiv \left| s_1-s_2\right| $. Following the procedure and
notation of Eq.\ (\ref{P2SumInt}), we can approximate the double
sum over $s_{1,2}$ by an integral over $r_{12}$ and obtain 
\begin{eqnarray}
\text{var}\left[ \frac{L\left( S_u\right) }{A\left( S\right) }\right]
&\approx &\left( \frac 4\Delta \right) ^2\frac 1{\pi r_{\max }^2}%
\int_0^{r_{\max }}\left[ P_2\left( u;r\right) -2P_3\left( u;\Delta ,r\right)
+P_4\left( u;\Delta ,r\right) \right] 2\pi rdr  \nonumber \\
&&-\left( \frac 4\Delta \right) ^2\left[ P_1\left( u\right) -P_2\left(
u;\Delta \right) \right] ^2,  \label{VarLAns}
\end{eqnarray}
where we have also replaced $n_sL_s/A_1$ by $4/\Delta $. By arguments
similar to those at the end of the previous subsection, the variance is
again inversely proportional to the total area $A\left( S\right) $ of the
observed region. The variance of $L$ will be considered in
more detail for Gaussian fields in Sec.~\ref{Sec:Gaussian}, where we
show that the precision of the approximations made in Eqs.\
(\ref{P3Approx})--(\ref{P4Approx}) is sufficient to evaluate the leading
term of the variance.

\subsection{The Euler characteristic}

We can use Eq.\ (\ref{MSum}) to calculate the Euler characteristic, but the
expression is considerably more complicated because the summation is
performed over all pairs, triples, etc.\ of pixels that have at least one
common point (since the Euler characteristic of any non-empty simply
connected set is equal to $1$). For the square lattice, up to four pixels
may have a common point, whereas for the hexagonal lattice, at most three
pixels intersect. A straightforward application of Eq.\ (\ref{MSum}) for a
hexagonal lattice gives 
\begin{equation}
\chi \left( S_u\right) =\sum_{s\in S}N_u\left( s\right) -\sum_{s_1\cap
s_2\neq \emptyset }N_u\left( s_1\right) N_u\left( s_2\right) +\sum_{s_1\cap
s_2\cap s_3\neq \emptyset }N_u\left( s_1\right) N_u\left( s_2\right)
N_u\left( s_3\right),  \label{ChiSum3}
\end{equation}
where the sums are taken over the sets of pixels which have at least one
common point, that is, over the sets of 2 and 3 adjacent pixels.
The corresponding expression for the square lattice is more complicated:
\begin{eqnarray}
\chi_1 \left( S_u\right) &=&\sum_{s\in S}N_u\left( s\right) -\sum_{s_1\cap
s_2\neq \emptyset }N_u\left( s_1\right) N_u\left( s_2\right) +\sum_{s_1\cap
s_2\cap s_3\neq \emptyset }N_u\left( s_1\right) N_u\left( s_2\right)
N_u\left( s_3\right)  \nonumber \\
&&-\sum_{s_1\cap s_2\cap s_3\cap s_4\neq \emptyset }N_u\left( s_1\right)
N_u\left( s_2\right) N_u\left( s_3\right) N_u\left( s_4\right) ,
\label{ChiSum4A}
\end{eqnarray}
where the sum now also includes contributions from four adjacent
pixels meeting at one vertex, along with both immediately and
diagonally adjacent pairs. The subscript 1 is used to distinguish this
definition from a slightly different one suggested in
Refs.\ (Adler 1981, Coles 1988). 
Namely, instead of using the union of all pixels $s$ with $N_u\left(
s\right) =1$, one could take the set formed by the {\em centers} of those
pixels and the lines connecting adjacent pixels (see Figs.\ 1a,b). As far as
the topology is concerned, the only difference is that the ``diagonally''
adjacent pixels are now considered disconnected. For lattice sizes much
smaller than the typical detail of the map, the Euler characteristic
calculated for this definition of $S_u$ would be equal to that for the
original definition, since in that case the diagonally adjacent pixels do
not affect the topology of the set. The Euler characteristic of the set $S_u$
is then given by 
\begin{equation}
\chi_2\left( S_u\right) =\sum_{s\in S}N_u\left( s\right)
-\sum_{L\left[ s_1\cap s_2\right] \neq 0}N_u\left( s_1\right) N_u\left(
s_2\right) +\sum_{s_1\cap s_2\cap s_3\cap s_4\neq \emptyset }N_u\left(
s_1\right) N_u\left( s_2\right) N_u\left( s_3\right) N_u\left( s_4\right) ,
\label{ChiSum4}
\end{equation}
with the summations going over sets of $1$, $2$, and $4$\ adjacent pixels,
without counting the diagonally adjacent pairs. (The third term in Eq.\ (\ref
{ChiSum4}) does not follow from Eq.\ (\ref{MSum}) but rather is added by
hand to compensate for the spurious ``holes'' inside the region hatched in
Fig.\ 1b.)

\begin{figure}[ht]
\centerline{\psfig{file=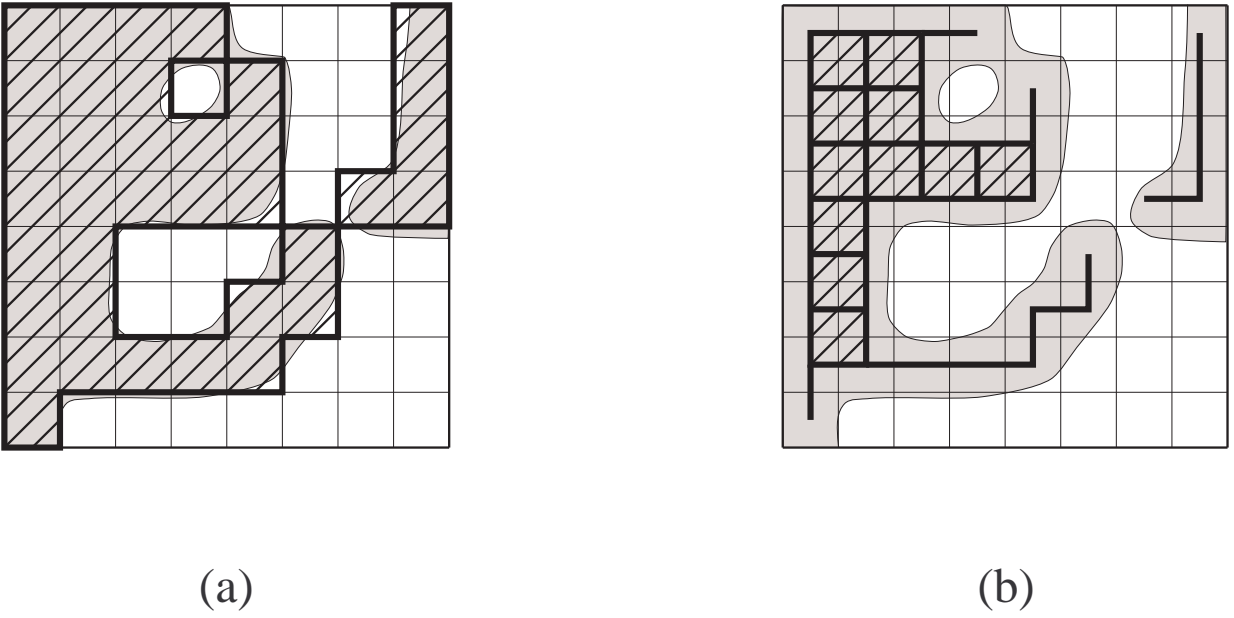,width=7in}}
\bigskip
\caption{
Two definitions of the Euler characteristic of a random field on a
lattice. The excursion set $S_u$ (the shaded curved region in both figures)
consists of all points where the field values are above the level $u$. (a)
The diagonally adjacent pixels are considered to be connected, making one
connected patch with two holes (shown in thick lines and hatching); the
Euler characteristic is $1-2=-1$. (b) The diagonally adjacent pixels are not
considered to be connected, which in effect reduces the original set $S_u$
to a smaller set shown in thick lines and hatching. The Euler characteristic
is equal to $2$. The difference between the two definitions disappears
with small enough lattice size.}
\label{Fig:1}
\end{figure}

The two definitions $\chi_1$ and $\chi_2$ 
on the square lattice are equally valid approximations of the
Euler characteristic of the
continuous excursion set, as will be any linear combination of the two.
By inspection of Eqs.\ (\ref{ChiSum4A}), (\ref{ChiSum4}) one finds that the
simple average of $\chi_1$ and $\chi_2$, 
\begin{equation}
\chi\left( S_u\right) =\frac{\chi_1 \left( S_u\right) +\chi_2\left(
S_u\right) }2,  \label{ChiBest}
\end{equation}
does not contain the sum over groups of four adjacent pixels and thus is the
simplest to deal with. This definition counts one-half of all diagonally
adjacent pairs of pixels. 
The averaging recipe has further advantages
when analyzing the Gaussian case,
suggesting that the definition (\ref{ChiBest}) is the
most suitable for square lattices.

Eqs.\ (\ref{ChiSum3}), (\ref{ChiSum4A})--(\ref{ChiBest}) can be used to
compute the Euler characteristic for a given map on a hexagonal and a square
lattice, respectively. As in Eq.\ (\ref{LSum0}), the sums over groups of
adjacent pixels can be transformed into the sums over neighbors of each
pixel, arriving to expressions which are manifestly linear in the total
number of pixels.

We now express the expectation value of the Euler characteristic through the
distributions $P_n$. {}From Eq.\ (\ref{ChiSum4}) for the square lattice we
obtain 
\begin{eqnarray}
\left\langle \chi_2 \left( S_u\right) \right\rangle &=&\sum_{s\in
S}\left\langle N_u\left( s\right) \right\rangle -\sum_{L\left[ s_1\cap
s_2\right] \neq 0}\left\langle N_u\left( s_1\right) N_u\left( s_2\right)
\right\rangle  \nonumber \\
&&\ +\sum_{s_1\cap s_2\cap s_3\cap s_4\neq \emptyset }\left\langle N_u\left(
s_1\right) N_u\left( s_2\right) N_u\left( s_3\right) N_u\left( s_4\right)
\right\rangle .  \label{ChiSum}
\end{eqnarray}
The first term of Eq.\ (\ref{ChiSum}) has already been calculated: 
\begin{equation}
\sum_{s\in S}\left\langle N_u\left( s\right) \right\rangle =NP_1\left(
u\right) . 
\end{equation}
In the second term, we only need to count the pairs of squares separated by
the distance $\Delta $. For each square there are four possible pairs, and
we have to divide by $2$ to prevent double counting of pairs. Therefore, the
second sum in Eq.\ (\ref{ChiSum}) is 
\begin{equation}
\sum_{L\left[ s_1\cap s_2\right] \neq 0}\left\langle N_u\left( s_1\right)
N_u\left( s_2\right) \right\rangle =2NP_2\left( u;\Delta \right) .
\label{NSum2}
\end{equation}
Finally, the third sum in Eq.\ (\ref{ChiSum}) is 
\begin{equation}
\sum_{s_1\cap s_2\cap s_3\cap s_4\neq \emptyset }\left\langle N_u\left(
s_1\right) N_u\left( s_2\right) N_u\left( s_3\right) N_u\left( s_4\right)
\right\rangle =NP_4\left( u;\Delta \right) ,  \label{NSum4}
\end{equation}
where $P_4\left( u;\Delta \right) $ is the probability that $f\left(
s\right) >u$ at four points $s_1$, $s_2$, $s_3$, $s_4$ situated at the
vertices of a square with side $\Delta $.
If we know the distributions $P_i$, $i=1,2,4$, for the random field $f$,
then the average Euler characteristic $\chi $ per unit area from Eq.~(\ref
{ChiSum}) is 
\begin{equation}
\frac{\left\langle \chi_2 \left( S_u\right) \right\rangle }{A\left(
S\right) }=%
\frac 1{\Delta ^2}\left[ P_1\left( u\right) -2P_2\left( u;\Delta \right)
+P_4\left( u;\Delta \right) \right] .  \label{ChiSquare}
\end{equation}

The other definition (\ref{ChiSum4A}) of the Euler characteristic for the
square lattice leads to the more complicated expression 
\begin{equation}
\frac{\left\langle \chi_1\left( S_u\right) \right\rangle }{A\left(
S\right) }=\frac 1{\Delta ^2}\left[ P_1\left( u\right) -2P_2\left( u;\Delta
\right) -2P_2\left( u;\Delta \sqrt{2}\right) +4P_3\left( u;\Delta ,\Delta 
\sqrt{2}\right) -P_4\left( u;\Delta \right) \right] ,  \label{ChiSquareA}
\end{equation}
where $P_3\left( u;\Delta ,\Delta \sqrt{2}\right) $ is the probability for
values at the vertices of a triangle with sides $\Delta $, $\Delta $, and $%
\Delta \sqrt{2}$ to be greater than $u$. Finally, the averaged
definition of $\chi $, Eq.~(\ref{ChiBest}), is expressed as 
\begin{equation}
\frac{\left\langle \chi\left( S_u\right) \right\rangle }{A\left( S\right) 
}=\frac 1{\Delta ^2}\left[ P_1\left( u\right) -2P_2\left( u;\Delta \right)
-P_2\left( u;\Delta \sqrt{2}\right) +2P_3\left( u;\Delta ,\Delta \sqrt{2}%
\right) \right] .  \label{Chi0Square}
\end{equation}

Analogous considerations for a regular hexagonal lattice yield the following
expression for the average Euler characteristic per unit area: 
\begin{equation}
\frac{\left\langle \chi \left( S_u\right) \right\rangle }{A\left( S\right) }=%
\frac 2{\Delta ^2\sqrt{3}}\left[ P_1\left( u\right) -3P_2\left( u;\Delta
\right) +2P_3\left( u;\Delta \right) \right] ,  \label{ChiHex}
\end{equation}
where $P_3\left( u;\Delta \right) $ is the probability for the field values
to be greater than $u$ at three points situated at vertices of an
equilateral triangle with side $\Delta $.

The variance of $\chi $ can be
calculated in essentially the same manner as the variance of $L$ in Eq.\ (%
\ref{VarLAns}), with a similar if more cumbersome expression as the result.
We omit that expression here and return to calculations of the
variance of $\chi $ in Sec.~\ref{Sec:Gaussian}.

\subsection{Boundary corrections}

A typical experimentally obtained temperature map does not cover the full
sky. Moreover, certain areas of the sky may be excluded from the map because
of the presence of foreground sources such as our Galaxy or other reasons.
The exclusions can be represented by taking the intersection of the full sky
with an appropriate ``window'' domain $W$. If we denote by $S_u$ the
excursion set of the temperature field if it were known throughout the full
sky, then Eqs.\ (\ref{AreaN}), (\ref{LSum}), (\ref{ChiSum3})--(\ref{ChiSum4}%
) of the previous subsections will yield the values of the Minkowski
functionals on the intersection of the full excursion set $S_u$ with the
window $W$ normalized to the area of $W$, i.e.\ one would obtain $M_i\left(
S_u\cap W\right) /A\left( W\right) $. Although these values can be regarded
as samples of the ``true'' functionals $M_i\left( S_u\right) /A(S)$
normalized to the full sky area $A(S)$, the existence of the boundary of $W$
introduces systematic errors which should be removed before comparing the
experimental values of $M_i\left( S_u\right) /A(S)$ with theoretical
predictions. This is essentially what we mean by boundary corrections.

In practice, we shall not need to assume that $S_u$ is defined on a full sky 
$S$, but only that $S$ covers a sky region large enough to disregard any
boundary effects. Since a full-sky map is defined on a sphere, in principle
additional complications due to curvature arise in calculations of the
Minkowski functionals; one indication of trouble is the fact that a sphere
cannot be tiled with an arbitrarily fine-grained lattice of regular
polygons. However, any curvature corrections should be negligible as long as
the pixel dimensions are small compared to the curvature scale, and any
lattice irregularities can be dealt with individually using the general
expressions Eqs.~(\ref{MSum}) and (\ref{VarM}). For simplicity, we limit the
following considerations to a suitably large flat region of the sky covered
by a regular lattice.

Some recipes for the boundary corrections have been proposed in the
literature, in particular, for calculating the Euler characteristic (Coles
1988). One natural procedure (Mecke et al. 1994) is based on the kinematic
property, Eq.~(\ref{kinematic}). Let $K_2=W$ be a fixed-shape window domain
and $K_1=S$ the underlying sky domain. By considering the measured $%
M_i(S_u\cap W)$ to be estimators for $\left\langle M_i(S_u\cap
W)\right\rangle $, Eq.~(\ref{kinematic}) can be inverted to obtain the
boundary correction formulas 
\begin{mathletters}
\begin{eqnarray}
\frac{A\left( S_u\right) }{A(S)} &=&\frac{A\left( S_u\cap W\right) }{A(W)},
\label{ACorr} \\
\frac{L\left( S_u\right) }{A(S)} &=&\frac{L\left( S_u\cap W\right) }{A(W)}-%
\frac{A\left( S_u\right) }{A(S)}\frac{L(W)}{A(W)},  \label{LCorr} \\
\frac{\chi \left( S_u\right) }{A(S)} &=&\frac{\chi \left( S_u\cap W\right) }{%
A(W)}-\frac{A\left( S_u\right) }{A(S)}\frac{\chi (W)}{A(W)}-\frac 1{2\pi }%
\frac{L\left( S_u\right) }{A(S)}\frac{L(W)}{A(W)}.  \label{ChiCorr}
\end{eqnarray}
(note that we need to substitute $L\left( S_u\right) $ from Eq.\ (\ref{LCorr}%
) to Eq.\ (\ref{ChiCorr})). These formulas hold for any shape of the window
domain $W$. It is of course impossible to calculate the Minkowski
functionals in a larger region than that in which the field is actually
measured, but if the measured region accurately represents the properties of
the field in the entire region, the above boundary correction formulas will
be accurate. Unless the cosmological microwave signal exhibits strongly
non-Gaussian features on the characteristic scales of the window domain
(i.e. individual pixels from point sources or the galactic cut) this is
likely to be an excellent assumption.

\section{Minkowski functionals for Gaussian fields}

\label{Sec:Gaussian}

In this section, we derive the expectation values of the Minkowski
functionals for Gaussian fields defined on a lattice. First, we treat the
area functional and compute its expectation value and variance. We then
explore the feature of smoothness of random fields which is important for
understanding the dependence of the Minkowski functionals on the pixel size.
We derive the expectation value of the boundary length functional on
arbitrary lattices and show explicitly its dependence on the lattice size
and geometry. The expectation value of the Euler characteristic is then
computed for the hexagonal lattice.

\subsection{Description of Gaussian fields}

For a homogeneous Gaussian field $f$ with zero mean, the two-point
correlation function 
\end{mathletters}
\begin{equation}
C\left( s_1,s_2\right) =\left\langle f\left( s_1\right) f\left( s_2\right)
\right\rangle \equiv C\left( r\right) ,\quad r\equiv \left| s_1-s_2\right| ,
\end{equation}
completely characterizes the field. Customarily, Gaussian random fields are
specified by their power spectrum $S\left( k\right) $ which is the Fourier
transform of the correlation function, expressed in two dimensions as 
\begin{equation}
C\left( r\right) =\int_0^\infty kJ_0\left( kr\right) S\left( k\right) dk.
\label{PowSpecDef}
\end{equation}
However, our present considerations are based on the physical space ($r$)
rather than on the momentum space ($k$), so the correlation function $%
C\left( r\right) $ is more relevant in this context than the power spectrum.

For simplicity, we assume that the variance of the field (which is the same
at all points) is $\left\langle f^2\right\rangle \equiv C\left( 0\right) =1$%
. (This can always be achieved by normalizing the field, $f=\bar{f}\sigma $,
if the original field had variance $\sigma ^2$. To recover the original
variables, one only needs to divide the function values $f$ by $\sigma $ and
the correlation function $C$ by $\sigma ^2$ in all formulas.) The
correlation function of the normalized field will then satisfy $\left|
C\left( r\right) \right| \leq 1$. Also, since correlations between distant
points should be absent, the correlation function must satisfy $C\left(
r=\infty \right) =0$.

Calculations of the Minkowski functionals require the
distribution functions $P_1\left( u\right) $, $P_2\left( u;r\right) $, and
so on, introduced in the previous section. These functions describe the
probabilities of encountering field values smaller than the level $u$
simultaneously at some given points $s_1$, $s_2$, ..., $s_n$ and can be
expressed through the $n$-point probability densities $p_n\left(
f_1,...,f_n\right) $ for values $f_i\equiv f\left( s_i\right) $ at these
points, cf.\ Eq.\ (\ref{PNDef}): 
\begin{equation}
P_{n\,}\left( u\right) =\int_u^\infty df_1...\int_u^\infty df_np_n\left(
f_1,...,f_n\right) .  \label{PthrupN}
\end{equation}
Since the field $f$ is Gaussian, all its $n$-point densities are
finite-dimensional Gaussian distributions with probability densities of the
form 
\begin{equation}
p_n\left( f_1,...,f_n\right) =\frac 1{\left( 2\pi \right) ^{n/2}\sqrt{\det B}%
}\exp \left( -B^{ij}f_if_j\right)
\end{equation}
with suitable coefficients $B^{ij}$ forming a positive-definite matrix. As
is well known, the matrix $B^{ij}$ is the inverse of the correlation matrix $%
C_{ij}\equiv \left\langle f\left( s_i\right) f\left( s_j\right)
\right\rangle $. Since the field is homogeneous, the coefficients $C_{ij}$
are completely determined by the separations between points $s_i$ and $s_j$,
i.e.\ $C_{ij}=C\left( \left| s_i-s_j\right| \right) $. Note that the
presence of Gaussian pixel noise can be easily described at this stage.
Imposition of noise with fixed standard deviation $\sigma ^2$ on a Gaussian
field with given $C\left( r\right) $ increases the value $C\left( 0\right) $
by $\sigma ^2$ while not changing $C\left( r\right) $ at any other point.
(For a discrete map, the function $C\left( r\right) $ is only available on a
discrete set of values $r=0$, $\Delta $, and so on.) We give the explicit
form of $p_2\left( f_1,f_2\right) $ in
Appendix \ref{App:P1}, including the pixel noise correction.

\subsection{Area and its variance}

We now consider the area functional of a Gaussian field. As follows from
Eq.\ (\ref{NormArea}), the expectation value of the normalized area
functional is equal to the value of the distribution function $P_1\left(
u\right) $ defined in the preceding section. The one-point probability
density for the values of the field is 
\begin{equation}
p_1\left( f\right) =\frac 1{\sqrt{2\pi }}\exp \left( -\frac{f^2}2\right) ,
\end{equation}
and therefore the one-point distribution function is 
\begin{equation}
P_1\left( u\right) =\int_u^\infty p_1\left( f\right) df=\frac 12\left( 1-%
\mathop{\rm erf}
\frac u{\sqrt{2}}\right) .  \label{P1Erf}
\end{equation}
This is the expectation value of the area of the excursion set $S_u$ per
unit area of the region.

The variance of the area functional is approximately given by Eq.\ (\ref
{VarArea3}), 
\begin{equation}
\text{var}\left[ \frac{A\left( S_u\right) }{A\left( S\right) }\right] =\frac 
1{A\left( S\right) }\int_0^{r_0}2\pi rdr\left\{ P_2\left( u;r\right) -\left[
P_1\left( u\right) \right] ^2\right\} .  \label{VarArea4}
\end{equation}
Since the integrand is assumed to vanish at $r>r_0$, the upper limit in Eq.\
(\ref{VarArea4}) can be extended to infinity. As shown in Eq.\ (\ref{P2I0})
of Appendix \ref{App:P2}, the two-point distribution function $P_2\left(
u;r\right) $ can be expressed through the correlation function $C\left(
r\right) $ as 
\begin{equation}
P_2\left( u;r\right) =\left[ P_1\left( u\right) \right] ^2+\int_0^{C\left(
r\right) }\frac{dC}{2\pi \sqrt{1-C^2}}\exp \left( -\frac{u^2}{1+C}\right) .
\label{P2fromApp}
\end{equation}
Then we can transform Eq.\ (\ref{VarArea4}) into 
\begin{eqnarray}
\text{var}\left[ \frac{A\left( S_u\right) }{A\left( S\right) }\right] &=&%
\frac 1{A\left( S\right) }\int_0^\infty 2\pi rdr\int_0^{C\left( r\right) }%
\frac{dC}{2\pi \sqrt{1-C^2}}\exp \left( -\frac{u^2}{1+C}\right)  \nonumber \\
&=&\frac 1{A\left( S\right) }\int_0^1\frac{r^2\left( C\right) dC}{2\sqrt{%
1-C^2}}\exp \left( -\frac{u^2}{1+C}\right) ,  \label{VarArea5}
\end{eqnarray}
where we have changed the order of integration and used the inverse $r\left(
C\right) $ of the correlation function $C\left( r\right) $. Thus, Eq.\ (\ref
{VarArea5}) allows one to calculate (at least numerically) the variance of
the area functional at a given level $u$ for any given correlation function $%
C\left( r\right) $. The maximum of the variance (at $u=0$), as can be seen
from Eq.\ (\ref{VarArea5}) and as we have checked for some sample functions $%
C\left( r\right) $, is proportional to $r_0^2/A\left( S\right) $, where $r_0$
is the typical distance at which correlations vanish, in agreement with the
qualitative estimate (\ref{VarAreaQ}).

We can also obtain the asymptotic form of the
variance in the limit of large levels $\left| u\right| $, for which the
integral in Eq.\ (\ref{VarArea5}) is dominated by the neighborhood of $C=1$
(small distances). We assume a Taylor expansion of the
correlation function $C\left( r\right) $ at small distances,
\begin{equation}
C\left( \Delta \right) =1 -C_1\left( \frac \Delta {r_1}%
\right) ^p+{\cal O}\left( \Delta ^{p+1}\right) ,  \label{PowerCor}
\end{equation}
where $C_1$, $r_1$ and $p<2$ are some constants. Then by expanding the
integrand in $\left( 1-C\right) $ we obtain the leading term
\begin{equation}
\text{var}\left[ \frac{A\left( S_u\right) }{A\left( S\right) }\right] 
\simeq
{\pi r_1^2\over A\left( S\right)C_1^{2/p}}\,\Gamma \left({4+p}\over{2p}\right)
\left({ u^2\over 2}\right)^{-(4+p)/2p}\exp \left( -{u^2\over 2}\right) .
\label{VarAAns}
\end{equation}
The square root of the variance (\ref{VarAAns}) is proportional
to $\exp \left( -u^2/4\right) $ and at large enough $u$ becomes large
relative to the ``signal'' itself which, as follows from Eq.\ (\ref{P1Erf}),
is proportional to $\exp \left( -u^2/2\right) $. This is to be expected for
all Minkowski functionals, since the signal at large $u$ comes from very few
map pixels. Thus, we can only use features of the Minkowski
functional profiles at levels $u$ within a few standard deviations from the
average.

\subsection{Smooth random fields}

A random field is {\em continuous} if the values at nearby points are highly
correlated. For a continuous field, the limit of the correlation function at
vanishing distance must be 
\begin{equation}
\lim_{\Delta \rightarrow 0}C\left( \Delta \right) =C\left( 0\right) \equiv 1,
\end{equation}
i.e.\ the correlation function must be continuous at $\Delta =0$. We shall
call a continuous random field {\em smooth} (with a somewhat relaxed rigor)
if its spatial derivatives are finite, i.e.\ have finite variances. To see
what this means mathematically, consider a homogeneous one-dimensional
random field $f\left( x\right) $ with a given correlation function $C\left(
r\right) $. The variance of a finite-differenced derivative $\left( f\left(
x+\Delta \right) -f\left( x\right) \right) /\Delta $ is 
\begin{equation}
\frac 1{\Delta ^2}\left\langle \left[ f\left( x+\Delta \right) -f\left(
x\right) \right] ^2\right\rangle =2\frac{1 -C\left( \Delta
\right) }{\Delta ^2}.  \label{VarDer}
\end{equation}
We call the field $f$ smooth if Eq.\ (\ref{VarDer}) has a finite limit at $%
\Delta \rightarrow 0$. This happens for instance if the correlation function
is regular at $\Delta =0$ with vanishing first derivative, 
\begin{equation}
C\left( \Delta \right) =1 -{1\over 2}C''(0)
\Delta ^2+{\cal O}\left( \Delta ^3\right) .  \label{SmoothCor}
\end{equation}
In general, if the asymptotic form of the correlation function for small $%
\Delta $ is a power law, Eq.~(\ref{PowerCor}),
then if $p<2$ the field is {\em not}
smooth. Physically, a random field is smooth if its small-wavelength modes
are suppressed strongly enough so that the field values at nearby points are
highly correlated. One can show, for instance, that if the small-wavelength
modes in the power spectrum of a Gaussian random field (cf.\thinspace Eq.\ (%
\ref{PowSpecDef})) are exponentially suppressed, as in the power spectrum 
\begin{equation}
S\left( k\right) \propto k^n\exp \left( -ak\right) \text{ as }k\rightarrow
+\infty ,
\end{equation}
then the correlation function $C\left( r\right) $ satisfies Eq.\ (\ref
{SmoothCor}) and the field is smooth.

Although the CMB temperature is expected to be smooth because of Silk
damping of CMB fluctuations on scales below a few arcminutes, the behavior
of the correlation function on larger scales, particularly the pixel size of
a given experimental map, may differ from Eq.\ (\ref{SmoothCor}). This could
make the CMB temperature field effectively non-smooth on the pixelization
scale. As we show below, excursion sets of non-smooth fields have a fractal
shape manifested by the scaling of the $L$ and $\chi$  functionals
with the pixel size. In our subsequent analysis, we shall {\em not} assume
that the correlation function possesses an expansion of type (\ref{SmoothCor}%
) on the relevant pixel scales.

\subsection{Boundary length for a Gaussian field}

The average boundary length $L\left( S_u\right) $ per unit area can be found
using Eq.\ (\ref{LAns}). We need to know the probability distribution $%
P_2\left( u;\Delta \right) $ for values $f_{1,2}$ at two points $s_{1,2}$
separated by the pixel size $\Delta $. Substituting Eq.\ (\ref{P1MinusP2})
from Appendix \ref{App:P2} into Eq.\ (\ref{LAns}), we obtain 
\begin{equation}
\frac{\left\langle L\left( S_u\right) \right\rangle }{A\left( S\right) }=%
\frac 2{\pi \Delta }\int_{C\left( \Delta \right) }^1\frac{dC}{\sqrt{1-C^2}}%
\exp \left( -\frac{u^2}{1+C}\right) .  \label{LInt}
\end{equation}
For small enough $\Delta $ the values of the correlation function $C\left(
\Delta \right) $ are close to $1$, so we can expand the above integral in
powers of the small parameter $\epsilon \equiv 1-C\left( \Delta \right) $
and obtain 
\begin{eqnarray}
\frac{\left\langle L\left( S_u\right) \right\rangle }{A\left( S\right) }&=&
\frac{2\sqrt{2\epsilon }}{\pi \Delta }\exp \left( -\frac{u^2}2\right)
\left[ 1+\frac{1-u^2}{12}\epsilon +\frac{3-6u^2+u^4}{160}\epsilon ^2\right.
\nonumber \\
&&\qquad +\frac{15-45u^2+15u^4-u^6}{2688}\epsilon ^3+ \left. \frac{%
105-420u^2+210u^4-28u^6+u^8}{55296}\epsilon ^4+{\cal O}
\left( \epsilon ^5\right)\right] .  \label{LSeries}
\end{eqnarray}
This expansion holds, as does Eq.\ (\ref{LAns}), for all regular lattices;
the expansion parameter $\epsilon $ is typically a power of the lattice step 
$\Delta $. We note that the dependence on $\Delta $ appears already in the
leading term. If the field is smooth and its correlation function is
described by Eq.\ (\ref{SmoothCor}), then it is easily seen that the leading
term of Eq.\ (\ref{LSeries}) is independent of $\Delta $ in the limit of
small $\Delta $. However, for non-smooth fields the dependence on $\Delta $
does not go away and may lead to formal divergence of $L$ at small $\Delta $.
For instance, if the correlation function satisfies Eq.\ (\ref{PowerCor})
in some range of $\Delta $, then the
leading term scales with $\Delta $ as $\Delta ^{p/2-1}$. The scaling
property of the leading term can be interpreted as a manifestation of
fractal geometry of the contour lines with fractal dimension $2-p/2>1$.

It is interesting to compare the series (\ref{LSeries}) and the result for
the length of level contours of a continuous smooth Gaussian field in two
dimensions [Adler 1981], 
\begin{equation}
\frac{\left\langle L\left( S_u\right) \right\rangle }{A\left( S\right) }=%
\frac{\sqrt{-C^{\prime \prime }\left( 0\right) }}2\exp \left( -\frac{u^2}2%
\right) .  \label{LAnsSmooth}
\end{equation}
With the correlation function (\ref{SmoothCor}), the leading term of Eq.\ (%
\ref{LSeries}) reproduces the continuous limit (\ref{LAnsSmooth}) up to the
factor $4/\pi $. This discrepancy is due to the fact that the length of a
discrete approximation of a line by a fixed regular lattice differs from the
length of the line by a geometric factor. This factor is easiest to derive
for the square lattice, where a curved line is approximated by horizontal
and vertical straight line segments. For lattice steps $\Delta $ much
smaller than the typical radius of curvature, the curve is locally
well-approximated by a straight line. If a straight line segment of length $%
L $ makes an angle $\alpha $ with the lattice (see Fig.\ 2a), then the
length of its lattice approximation is $\tilde{L}=L\left( \cos \alpha +\sin
\alpha \right) $. Averaging this over all angles $\alpha $ (it suffices to
consider $0<\alpha <\pi /2$), we obtain 
\begin{equation}
\left\langle \tilde{L}\right\rangle _\alpha =\frac 2\pi \int_0^{\pi
/2}L\left( \cos \alpha +\sin \alpha \right) d\alpha =\frac 4\pi L.
\end{equation}
This means that the lattice approximation of the length of a curve with
isotropically distributed tangent angle is wrong by a factor of $4/\pi $ on
the average. By a similar argument it can be shown that any tiling by
identical regular polygons gives rise to the same factor $4/\pi $. (In the
case of non-regular tilings, the correction factor will generally differ
from $4/\pi $, and also the higher terms of the series (\ref{LSeries}) will
be different.)

A more intuitive way to obtain this correction factor $4/\pi $ is to
consider a square with an inscribed circle. The length of the lattice
approximation to the circle is equal to the perimeter of the square (see
Fig.\ 2b). The ratio of the circumferences is $4/\pi $. The circle
effectively averages over all orientations of the line, therefore the result
also holds on the average for curved lines with isotropically distributed
tangent angles.

\begin{figure}[ht]
\centerline{\psfig{file=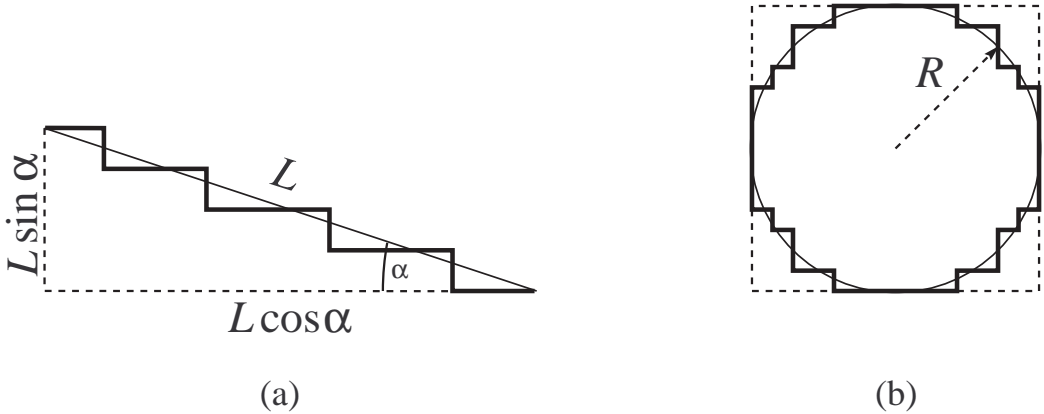,width=7in}}
\bigskip
\caption{Geometric correction factor for curves on a lattice. (a)
Approximation of a straight line segment $L$ making an angle $\alpha $ with
the lattice has length $L\left( \cos \alpha +\sin \alpha \right) $.
Averaging this over the angles $\alpha $ yields $4L/\pi $. (b) The
circumference of a circle, $2\pi R$, is approximated by a rectangular
lattice. The result, even for an arbitrarily fine lattice, is $8R$, which is
off by the same factor $4/\pi $.}
\label{Fig:2}
\end{figure}

Apart from the geometric correction factor, a comparison of the series (\ref
{LSeries}) with the continuous limit (\ref{LAnsSmooth}) shows that higher
terms of the series contain polynomials in $u$ which distort the Gaussian
profile. The resulting small deviation from Eq.\ (\ref{LAnsSmooth}) is a
direct consequence of pixelization. While Eq.\ (\ref{LSeries}) holds for any
regular lattice, in particular for square and hexagonal lattices, 
in asymmetric lattices the coefficients of the series for $L$
depend on the pixel geometry.

Application of the general formula of Eq.\ (\ref{VarLAns}) for the variance
to the Gaussian case requires calculation of the distribution $P_4\left(
u;\Delta ,r\right) $ for the two pairs of closest neighbor points separated
by the distance $r$. In Appendix \ref{App:P2}, an expansion of the
distribution $P_4\left( u;\Delta ,r\right) $ in powers of $\epsilon \equiv
1-C\left( \Delta \right) $ is found. We retain only the leading term of this
expansion, because the error of the approximation in Eq.\ (\ref{VarLAns}) is
typically worse than the contribution $O\left( \epsilon \right) $ of the
higher terms of the expansion. The resulting expression for the variance of $%
L$ is 
\begin{equation}
\text{var}\left[ \frac{L\left( S_u\right) }{A\left( S\right) }\right]
\approx \left( \frac 2{\pi \Delta }\right) ^2\frac{2\epsilon }{\pi r_{\max
}^2}\int_0^{r_{\max }}\left[\left(1-C(r)^2\right)^{-1/2}
\exp \left( -{u^2\over 1+C(r)}\right)
-\exp \left( -u^2\right)
\right] 2\pi rdr.  \label{VarLGAns}
\end{equation}
As in the previous section, one can transform this integral to the variable $%
C\left( r\right) $ and show that the asymptotic form of the variance (\ref
{VarLGAns}) at large $\left| u\right| $ is proportional to $\exp \left(
-u^2/2\right) $.

\subsection{The Euler characteristic}

Finally we consider the Euler characteristic of the excursion set per unit
area. We first consider the simpler case of a hexagonal lattice for which
the Euler characteristic is given by Eq.\ (\ref{ChiHex}). Using Eq.\ (\ref
{ChiComb}) from Appendix \ref{App:P2} and expanding it in powers of the
small parameter $\epsilon \equiv 1-C\left( \Delta \right) $, we obtain the
series 
\begin{eqnarray}
\frac{\left\langle \chi \left( S_u\right) \right\rangle }{A\left( S\right) }
&=&\frac{2u\exp (-u^2/2)}{\left( 2\pi \right) ^{3/2}\Delta ^2}\epsilon
\left[ 1+5\frac{3-u^2}{36}\epsilon +7\frac{15-10u^2+u^4}{540}\epsilon
^2\right.  \nonumber \\
&&+83\frac{105-105u^2+21u^4-u^6}{90720}\epsilon ^3 
\left. +319\frac{945-1260u^2+378u^4-36u^6+u^8}{6123600}\epsilon ^4+{\cal O}%
\left( \epsilon ^5\right) \right] .  \label{Chi3Series}
\end{eqnarray}
The leading term of the series (\ref{Chi3Series}) coincides with the formula
of Adler (1981) for the average Euler characteristic of excursion sets of a
smooth Gaussian field, 
\begin{equation}
\frac{\left\langle \chi \right\rangle }{A\left( S\right) }=
{|C''(0)|\over (2\pi)^{3/2}}\, u\exp \left(-{u^2\over 2}\right)
\label{ChiAdler}
\end{equation}

The case of a square lattice is more complicated. We can use the methods of
Appendix \ref{App:P2} to obtain a series expansion of Eqs.\ (\ref{ChiSquare}%
)--(\ref{Chi0Square}). However, since the point configurations for the
required distributions $P_3$ and $P_4$ contain neighbors separated by
distances $\Delta $ and $\Delta \sqrt{2}$, there are now two small
parameters, namely $\epsilon \equiv 1-C\left( \Delta \right) $ and $%
b\epsilon \equiv 1-C\left( \Delta \sqrt{2}\right) $, where we have
introduced the parameter $b$ which is of order $1$ and which characterizes
the behavior of the correlation function $C\left( r\right) $ at smallest
distances available on the lattice, 
\begin{equation}
b\equiv \frac{1-C\left( \Delta \sqrt{2}\right) }{1-C\left( \Delta \right) }.
\end{equation}
By definition, $b>0$; moreover, one can derive the condition $b<2$ from the
requirement of normalizability of the four-point distribution $p_4$ for the
corners of a closest neighbor square (cf.\ Appendix \ref{App:P1}, Eq.\ (\ref
{Multivariate}) with the correlation matrix of Eq.\ (\ref{CM4})). A smooth
Gaussian field satisfying Eq.\ (\ref{SmoothCor}) is described by $b\approx 2$%
, and the deviation of $b$ from $2$ is of order $1-C\left( \Delta \right) $.
{}From a given experimental map, the parameter $b$ can be estimated in linear
time. If the value of $b$ is significantly smaller than $2$, namely if $%
2-b\gg 1-C\left( \Delta \right) $, we conclude that the underlying
field is not smooth on scales $\Delta $.

After computing the expansions of Eqs.\ (\ref{ChiSquare})--(\ref{Chi0Square}%
) in $\epsilon $ (while keeping the parameter $b$ constant), one finds in
each case an expression similar to Eq.\ (\ref{Chi3Series}), except that both
the leading term and the coefficients of the polynomials in $u$ in the
higher terms depend on $b$. However, the most significant difference is that
the leading term of the expansion of Eqs.\ (\ref{ChiSquare}) and (\ref
{ChiSquareA}) is not of order $\epsilon $ but of order $\sqrt{\epsilon }$,
and also the expansion parameter of the series is $\sqrt{\epsilon }$ instead
of $\epsilon $. The first two terms of the expansion of Eq.\ (\ref
{ChiSquareA}) are 
\begin{equation}
\frac{\left\langle \chi_2\left( S_u\right) \right\rangle }{A\left(
S\right) }= {1\over\sqrt{2}\pi^2\Delta^2}\left(
2\arccos{b\over 4-b}-\sqrt{b}\arccos {3b-4\over 4-b}\right)
\exp\left(-{u^2\over 2}\right)\sqrt{\epsilon }
+{\sqrt{b(4-b)}\over(2\pi)^{3/2}\Delta^2}
\,u\exp\left(-{u^2\over2}\right)\epsilon 
+{\cal O}\left(\epsilon ^{3/2}\right) .  \label{Chi4ASeries}
\end{equation}
At $b=2-O\left( \epsilon \right) $, which corresponds to
a smooth field, the first term as well as all terms with non-integer powers
of $\epsilon $ vanish, and the series (\ref{Chi4ASeries}) becomes again a
series in $\epsilon $. For non-smooth fields, however, Eq.\ (\ref
{Chi4ASeries}) does not correspond to its continuous limit Eq.\ (\ref
{ChiAdler}).

A similar result is obtained for $\chi_1 $, Eq.\ (\ref
{ChiSquare}). The situation is however different with the Euler
characteristic defined by Eq.\ (\ref{Chi0Square}): all terms with
non-integer powers of $\epsilon $ cancel, and the result is 
\begin{eqnarray}
\frac{\left\langle \chi\left( S_u\right) \right\rangle }{A\left( S\right) 
} &=&{\sqrt{b(4-b)}\over(2\pi)^{3/2}\Delta^2}
\,u\exp\left(-{u^2\over 2}\right)\epsilon 
\left[ 1+\frac{\left( 3-u^2\right) \left( 4+2b-b^2\right) }{%
12\left( 4-b\right) }\epsilon \right. \nonumber\\
&&\qquad +\frac{\left( b^4-6b^3+6b^2+2b+18\right) \left(
u^4-10u^2+15\right) }{180\left( 4-b\right) ^2}\epsilon ^2 \nonumber\\
&&\qquad\left. +\,\frac{\left( b^6-10b^5+30b^4-20b^3-12b^2+8b-80\right)
\left( u^6-21u^4+105u^2-105\right) }{3360\left( 4-b\right) ^3}\epsilon
^3\right] +{\cal O}\left( \epsilon ^5\right)  \label{Chi40Series}
\end{eqnarray}
The series (\ref{Chi40Series}) is reduced to Eq.~(\ref{Chi3Series}) 
at $b=1$. 
The leading term of this series is, up to a constant, the same as the
continuous limit Eq.\ (\ref{ChiAdler}); the correspondence is exact at
$b=2$. For this reason, we suggest that the ``averaged'' definition of
$\chi$, Eq.~(\ref{Chi0Square}), be used for square lattice calculations.
As in the previous section, we note the scaling properties of Eqs.\ (\ref
{Chi3Series})--(\ref{Chi40Series}) with $\Delta $ and interpret them as a
manifestation of the fractal structure of the excursion set. For smooth
fields, the dependence on $\Delta $ cancels out, while for non-smooth fields
it may lead to a divergence of $\chi $ at small $\Delta $.

The variance of $\chi $ can be estimated similarly to the variance of the
boundary length $L$, except that the expressions of Eqs.\ (\ref{Chi0Square}%
), (\ref{ChiHex})\ for $\chi $ contain $3$-point distributions $P_3$ and
therefore its variance requires calculation of the $6$-point distribution $%
P_6$. This calculation can also be performed by methods presented in
Appendix \ref{App:P2} and, as in the case of the variance of $L$, only the
leading term of the resulting expansion suffices up to the precision of
approximation. We give the result for the variance of the Euler
characteristic only in the simplest case on hexagonal lattice: 
\begin{equation}
\text{var}\left[ \frac{\chi \left( S_u\right) }{A\left( S\right) }\right]
\simeq \frac{\epsilon ^2}{\Delta ^4\pi ^3r_{\max }^2}\int_0^{r_{\max
}}\left[ \frac{u^2\left(
1-C\right) +C\left( 1+C\right)}{\sqrt{1-C^2}\left( 1-C\right)
\left( 1+C\right) ^2}\exp \left( -\frac{u^2}{1+C}\right)
-u^2\exp \left( -u^2\right) \right] rdr,
\end{equation}
where we imply $C\equiv C\left( r\right) $. The properties of the variance
of $\chi $ are similar to those of the variances of the other two Minkowski
functionals, as has been discussed above.

\section{SUMMARY AND DISCUSSION}

In this paper, we have proposed the Minkowski functionals -- area,
boundary length, and Euler characteristic -- of excursion sets as a
probe of a map's Gaussianity. They are all linear functions of the
number of map pixels, and thus are easy to compute from a
map. Gaussian pixel noise is straightforward to include, and irregular
map boundaries, arising from partial sky coverage or cuts of the data,
can be accounted for in a natural manner. Finally, for Gaussian
distributions the functionals can be calculated exactly and their
variances estimated given the correlation function of the
distribution, making a test of
Gaussianity straightforward and eliminating the need for Monte Carlo
simulations.

We have presented a complete formulation of Minkowski functionals on a
two-dimensional pixelized map, including explicit boundary corrections
and pixel-size dependences. We propose an alternative definition of
the Euler characteristic for pixelized maps which possesses nice
calculational properties compared with previously used definitions.
In addition, we provide explicit forms of the functionals and their
variances for
the case of Gaussian distributions, including pixelization
corrections. 

Minkowski functionals are simplest to calculate 
for maps with hexagonal pixelation schemes, because no more than three
pixels can ever adjoin a single point and all adjoining pixels are the
same distance apart. This property also makes
estimation of the Euler functional more straightforward
and allows easier estimation of the variances of Gaussian map
functionals. Hexagonal pixelizations also have other nice properties
like optimal smoothness and regularity (Tegmark 1996). Analysis of
maps with square pixels will be a bit more technically involved, but
not fundamentally different. Actual full-sky maps will always have
irregular points in the pixelization lattice (since the sphere cannot
be tiled by identical polygons) which will need to be handled on an
individual basis. No conceptual difficulties arise in extending the
Minkowski functionals to a sphere as long as the pixelization scale is
small compared to the curvature scale, which will always be the case
for maps of moderately high resolution.

It is impossible to make a general statement about how well these
functionals will distinguish non-Gaussianity in a map because the
variety of non-Gaussianity is endless. Since Minkowski functionals are
ensemble or map averages, it is likely they will be best at
picking out non-Gaussian distributions which are spread over an entire
map, as opposed to isolated, sharp features which will be lost in the
averaging. Even if the primordial perturbations are
Gaussian-distributed (as expected in inflationary models), 
potentially interesting non-Gaussianity might arise from
weak gravitational lensing
(Bernardeau 1996) or from Sunyaev-Zeldovich distortions. For a
specific kind of non-Gaussian contribution, simulations of the
resulting sky will be required to determine the utility of Minkowski
functional analysis.

Any test of primordial non-Gaussianity in CMB maps is likely to be
challenging with real data. Just as noise correlations can mimic
CMB power (Dodelson and Kosowsky, 1995), any noise correlation will
create non-Gaussianity in a map (Kogut et al. 1994). 
A meaningful Gaussianity test
requires detailed understanding of an experiment's noise
properties. The best bet for Gaussianity tests are those experiments
with simple noise properties; the MAP satellite design, for example,
was driven largely by the desire for very simple noise
properties. Other unavoidable noise correlations are induced by
projecting out foreground contamination, but these correlations can
likely be understood in detail through modelling and simulations.
Conversely, for experiments with complicated noise properties,
Minkowski functional analysis may aid in characterizing the noise
as long as the underlying fluctuations are close to Gaussian.

Gaussian-distributed initial fluctuations are a decisive prediction of
inflationary models. Upcoming CMB data sets in the form of
high-resolution maps will allow meaningful tests of this
prediction. The theoretical framework presented in this paper refines
and extends the current menu of Gaussianity tests, and we anticipate
that Minkowski functionals will be a basic component of future
microwave background map analysis. 

\acknowledgments

S.W. thanks Thomas Buchert, Martin Kerscher, R\"udiger Kneissl, Jens
Schmalzing, and Roberto Trasarti-Battistoni for helpful
discussions and hospitality during his stay in Munich. This work has been
partially supported by the Society of Fellows at Harvard University.

\appendix

\section{Gaussian probability densities in several variables}

\label{App:P1}

We first consider the simplest case of the Gaussian two-point
probability density $p_2\left( x,y;\Delta \right) $ for field values $x$, $y$
at points $s_{1,2}$ separated by the distance $\Delta $. For a normalized
Gaussian field, the correlations are 
\begin{eqnarray*}
\left\langle x^2\right\rangle &=&\left\langle y^2\right\rangle =1, \\
\left\langle xy\right\rangle &=&C\left( \Delta \right) .
\end{eqnarray*}
We can easily diagonalize the correlation matrix if we notice that 
\begin{eqnarray*}
\left\langle \left( x+y\right) \left( x-y\right) \right\rangle &=&0, \\
\left\langle \left( x\pm y\right) ^2\right\rangle &=&2\left( 1\pm C\left(
\Delta \right) \right) ,
\end{eqnarray*}
which means that $\left( x+y\right) $ and $\left( x-y\right) $ are
independent Gaussian variables with known variances. One can then write the
joint distribution for $\left( x\pm y\right) $ as a product of two
Gaussians, 
\begin{equation}
p_2\left( x,y;\Delta \right) dxdy=\frac{dxdy}{2\pi \sqrt{1-C\left( \Delta
\right) ^2}}\exp \left[ -\frac 14\left( \frac{\left( x+y\right) ^2}{%
1+C\left( \Delta \right) }+\frac{\left( x-y\right) ^2}{1-C\left( \Delta
\right) }\right) \right] .  \label{Bivariate}
\end{equation}

More generally, a multivariate Gaussian distribution in $n$ variables $x_1$,
..., $x_n$ is completely specified by a symmetric correlation matrix $%
C_{ij}\equiv \left\langle x_ix_j\right\rangle $, and the joint probability
density is expressed through the inverse matrix $C^{-1}$ by 
\begin{equation}
p_n\left( {\bf x};C_{ij}\right) d^n{\bf x}=\frac{d^n{\bf x}}{\left( 2\pi
\right) ^{n/2}\sqrt{\det C_{ij}}}\exp \left[ -\frac 12%
\sum_{i,j=1}^nC_{ij}^{-1}x_ix_j\right] .  \label{PnCinv}
\end{equation}
If the matrix $C_{ij}$ is explicitly diagonalized and its eigenvalues $%
\lambda _i$ and the corresponding orthonormal eigenvectors $E_i^j$ are
known, then the probability density $p_n\left( {\bf x};C_{ij}\right) $ can
be written as a product of Gaussians similar to Eq.\ (\ref{Bivariate}), 
\begin{equation}
p_n\left( {\bf x};C_{ij}\right) =\frac 1{\left( 2\pi \right) ^{n/2}\sqrt{%
\det C_{ij}}}\exp \left[ -\frac 12\sum_{i=1}^n\frac{\left(
\sum_{j=1}^nE_i^jx_j\right) ^2}{\lambda _i}\right] .  \label{Multivariate}
\end{equation}
However, this explicit form of the distribution density will not be as
useful as its representation as a Fourier transform of its generating
function $\tilde{p}_n\left( {\bf k};C_{ij}\right) $, 
\begin{equation}
p_n\left( {\bf x};C_{ij}\right) \equiv \int \frac{d^n{\bf k}}{\left( 2\pi
\right) ^n}\tilde{p}_n\left( {\bf k};C_{ij}\right) =\int \frac{d^n{\bf k}}{%
\left( 2\pi \right) ^n}\exp \left[ i{\bf kx-}\frac 12%
\sum_{i,j=1}^nC_{ij}k_ik_j\right] .  \label{PnFourier}
\end{equation}
This form of the probability density will be used in Appendix \ref{App:P2}
for calculations of the probability integrals $P_n$ defined by Eq.\ (\ref
{PNDef}).

The distributions for values of a homogeneous and isotropic Gaussian field
at the vertices of an equilateral triangle or a square are found as special
cases of the multivariate Gaussian distribution. The three-point
distribution $p_3\left( x,y,z;C\right) $ is characterized by a single
parameter, the correlation $C\equiv C_{12}=\left\langle xy\right\rangle
=\left\langle yz\right\rangle =\left\langle zx\right\rangle $ between any
two different points, while $C_{ii}\equiv 1$. The correlation matrix $C_{ij}$
is 
\begin{equation}
C_{ij}=\left( 
\begin{array}{ccc}
1 & C & C \\ 
C & 1 & C \\ 
C & C & 1
\end{array}
\right) .  \label{CM3}
\end{equation}
Similarly, one obtains the four-point distribution $p_4\left( w,x,y,z\right) 
$ for values at the vertices of a square. This time the distribution has two
free parameters, namely the correlations between the adjacent points $%
C_{12}\equiv C$ and between the diagonally opposing points $C_{23}\equiv D$.
The correlation matrix $C_{ij}$ in that case is 
\begin{equation}
C_{ij}=\left( 
\begin{array}{cccc}
1 & C & C & D \\ 
C & 1 & D & C \\ 
C & D & 1 & C \\ 
D & C & C & 1
\end{array}
\right) .  \label{CM4}
\end{equation}
The explicit forms of the distributions $p_3\left( x,y,z;C\right) $ and $%
p_4\left( w,x,y,z;C,D\right) $ can be obtained using Eq.\ (\ref{Multivariate}%
), but we omit them as they will not be useful for our calculations.

Finally we briefly describe how a known level of Gaussian pixel noise
present in experimental maps can be represented in this formalism. If we
assume that Eq.~(\ref{Multivariate}) describes the $n$-point distribution
obtained from a noiseless map, then pixel noise with a fixed variance $%
\sigma ^2$ (in units of the field variance) will simply increase the
diagonal elements of the matrix $C_{ij}$ by $\sigma ^2$, while not changing
any other correlations. Thus pixel noise can be straightforwardly
incorporated by making the replacements $C_{ij}\rightarrow C_{ij}/(1+\sigma
^2)$ for $i\neq j$ in the correlation matrix.

\section{The Gaussian probability integrals}

\label{App:P2}

In this appendix we give some derivations for the distributions $P_n\left(
u\right) $ which define the probability for the values of a Gaussian random
field at $n$ given points to be above the threshold value $u$. Generally,
these distributions, as defined by Eq.\ (\ref{PthrupN}), involve multiple
integrals of Gaussians of the type (\ref{PnCinv}) and cannot be exactly
evaluated. However, if the correlation between the point values is high ($%
C_{ij}\approx 1$), as is the case for neighboring points of a fine lattice,
one can develop series expansions of $P_n\left( u;C_{ij}\right) $ in powers
of $\left( 1-C_{ij}\right) $. We follow the calculations presented in
Hamilton et al. (1986), where a method for computing the series expansions
for the distributions $P_2$, $P_3$, and $P_4$ was presented, the expansion
parameter being the lattice step $\Delta $. As we have seen in Sec.~\ref
{Sec:MinkLattice}, knowledge of these three distributions suffices to find
the three Minkowski functionals on the plane. However, we do not assume an
expansion of type (\ref{SmoothCor}) for the correlation function and keep $%
\left( 1-C_{ij}\right) $ as expansion parameters instead of the lattice step 
$\Delta $, so that our results are applicable not only to smooth Gaussian
fields. For completeness, we explain the method of calculation in some
detail; this will also enable us to clarify the limitations of this method
when applied to non-Gaussian distributions.

We start with the Fourier representation of a Gaussian two-point
symmetric probability density, 
\begin{eqnarray}
p_2\left( {\bf x};C\right) &=&\int \frac{d^2{\bf k}}{\left( 2\pi \right) ^2}%
\exp \left[ i{\bf kx-}\frac 12\left( k_1^2+k_2^2\right) -Ck_1k_2\right] 
\nonumber \\
&\equiv &\int \frac{d^2{\bf k}}{\left( 2\pi \right) ^2}e^{i{\bf kx}}\tilde{p}%
_2\left( {\bf k};C\right) .  \label{PnFourier1}
\end{eqnarray}
The distribution (\ref{PnFourier1}) is symmetric with respect to an
interchange of $x_{1,2}$ and is characterized by the mean value $%
\left\langle x_i\right\rangle =0$, variance $\left\langle x_i^2\right\rangle
=1$, and correlation $\left\langle x_1x_2\right\rangle =C$ with $\left|
C\right| \leq 1$. The probability integral $P_2\left( u;C\right) $ is
defined by 
\begin{equation}
P_2\left( u;C\right) \equiv \int_u^\infty dx\int_u^\infty dy\,p_2\left( {\bf %
x};C\right) .  \label{P2GenInt}
\end{equation}
This coincides with the distribution $P_2\left( u;C\left( \Delta \right)
\right) $ defined by Eq.\ (\ref{P2Def}).

One then takes the partial derivative of $P_2\left( u;C\right) $ with
respect to $C$. Since 
\begin{equation}
\frac \partial {\partial C}\tilde{p}_2\left( {\bf k};C\right) =-k_1k_2\tilde{%
p}_2\left( {\bf k};C\right) ,
\end{equation}
it follows that 
\begin{equation}
\frac \partial {\partial C}P_2\left( u;C\right) =-\int_u^\infty
dx\int_u^\infty dy\int \frac{d^2{\bf k}}{\left( 2\pi \right) ^2}k_1k_2e^{i%
{\bf kx}}\tilde{p}_2\left( {\bf k};C\right) .  \label{Interm1}
\end{equation}
To change the order of integration in Eq.\ (\ref{Interm1}),
it is necessary to replace the infinite upper limits in the
integrations over $x$ and $y$ by a finite quantity $M$ and take the limit of 
$M\rightarrow \infty $ afterwards. The integrations over $x$ and $y$ yield 
\begin{equation}
\int_u^Mdx\int_u^Mdye^{i{\bf kx}}=\frac{e^{ik_1M}-e^{ik_1u}}{ik_1}\frac{%
e^{ik_2M}-e^{ik_2u}}{ik_2}.
\end{equation}
The factor $k_1k_2$ in Eq.\ (\ref{Interm1}) cancels. The
distribution density $p_2\left( {\bf x};C\right) $ decays at spatial
infinity, so
\begin{equation}
\frac \partial {\partial C}P_2\left( u;C\right) =\lim_{M\rightarrow \infty
}\left[ p_2\left( u,u;C\right) -2p_2\left( u,M;C\right) +p_2\left(
M,M;C\right) \right] =p_2\left( u,u;C\right) .  \label{DCRel}
\end{equation}
By means of this relation, the double probability integral in Eq.\ (\ref
{P2GenInt}) is reduced to a single integral 
\begin{equation}
P_2\left( u;C\right) =P_2\left( u;1\right) -\int_C^1\,p_2\left( u,u;C\right)
dC.  \label{P2Rel}
\end{equation}
Here we choose the boundary condition at $C=1$ which
corresponds to the limit of two coincident points, where the two-point
distribution degenerates into a one-point Gaussian, 
\begin{equation}
p_2\left( x,y;1\right) =\frac{\delta \left( x-y\right) }{\sqrt{2\pi }}\exp
\left( -\frac{x^2}2\right) ,
\end{equation}
so it easily follows that 
\begin{equation}
P_2\left( u;1\right) \equiv P_1\left( u\right) =\frac 12\left( 
1-\mathop{\rm erf}
\left[ \frac u{\sqrt{2}}\right] \right) .
\end{equation}
Now substituting into Eq.\ (\ref{P2Rel}) the explicit form
of the distribution from Eq.\ (\ref{Bivariate}), 
\begin{equation}
p_2\left( u,u;C\right) =\frac 1{2\pi \sqrt{1-C^2}}\exp \left[ -\frac{u^2}{1+C%
}\right],
\end{equation}
gives 
\begin{equation}
P_2\left( u;C\right) =P_1\left( u\right) -\int_C^1\frac{dC}{2\pi \sqrt{1-C^2}%
}\exp \left[ -\frac{u^2}{1+C}\right] .  \label{P1MinusP2}
\end{equation}
Although the resulting integral cannot be evaluated analytically, it is
easily expanded in powers of $\left( 1-C\right) \equiv \epsilon $, 
\begin{eqnarray}
\int_{1-\epsilon }^1\frac{dx}{\sqrt{1-x^2}}\exp \left[ -\frac{u^2}{1+x}%
\right] =\sqrt{2\epsilon } &&\,\exp \left( -\frac{u^2}2\right) \left[ 1+%
\frac{1-u^2}{12}\epsilon +\frac{3-6u^2+u^4}{160}\epsilon ^2\right.  
\nonumber \\
+\frac{15-45u^2+15u^4-u^6}{2688}\epsilon ^3+ &&\left. \frac{%
105-420u^2+210u^4-28u^6+u^8}{55296}\epsilon ^4+O\left( \epsilon ^5\right)
\right] .  \label{IntSer2}
\end{eqnarray}

Another possible choice of the boundary condition is $C=0$.
The two-point distribution $p_2$
degenerates into a product of two one-point distributions $p_1$, which leads
to the following result,
\begin{equation}
P_2\left( u;C\right) =\left[ P_1\left( u\right) \right] ^2+\int_0^1\frac{dC}{%
2\pi \sqrt{1-C^2}}\exp \left[ -\frac{u^2}{1+C}\right] .  \label{P2I0}
\end{equation}

As has been already noted by Hamilton et al. (1986), an extension of this
technique to non-Gaussian distributions is problematic. Namely, one might
consider the Fourier representation of a non-Gaussian two-point distribution 
\begin{equation}
p_2\left( {\bf x};C,D,...\right) =\int \frac{d^2{\bf k}}{\left( 2\pi \right)
^2}\exp \left[ i{\bf kx-}\frac 12\left( k_1^2+k_2^2\right) -Ck_1k_2-\sum
D_{ijl}k_ik_jk_l+...\right] ,
\end{equation}
where the coefficients $D_{ijl}$ and the (omitted) higher-order terms of the
series in ${\bf k}$ represent deviations from Gaussianity. Then a formal
relation similar to Eq.\ (\ref{DCRel}) may be derived: 
\begin{equation}
\frac \partial {\partial C}P_2\left( u;C,D,...\right) =p_2\left(
u,u;C,D,...\right) . 
\end{equation}
However, the limit of $p_2$ at $C=1$ and fixed values of $D_{ijl}$ and
higher moments does not correspond to a limit of the two-point distribution
with coincident points, since in that limit not only the second moment $C$,
but also the higher moments $D_{ijl}$, ... are changed to make the
expression under the exponential a function only of $\left( k_1+k_2\right) $%
. Since $p_2\left( C=1,D,...\right) $ does not degenerate into a one-point
distribution as before, one cannot make use of Eq.\ (\ref{P2Rel}) to compute
the integral distribution $P_2$.

Having covered the derivation of the two-point distribution $P_2$ in detail,
we now sketch the more complicated cases of the distributions $P_3$
and $P_4$. The three-point distribution is
\begin{equation}
P_3\left( u;C\right) \equiv \int_u^\infty dx\int_u^\infty dy\int_u^\infty
dz\,p_3\left( x,y,z;C\right) 
\end{equation}
for a symmetric three-point distribution density $p_3$ corresponding to the
correlation matrix of Eq.\ (\ref{CM3}). The Fourier image of $p_3\left(
x,y,z;C\right) $ is 
\begin{equation}
\tilde{p}_3\left( k_1,k_2,k_3;C\right) =\exp \left[ -\frac 12\left(
k_1^2+k_2^2+k_3^2\right) -C\left( k_1k_2+k_2k_3+k_3k_1\right) \right] .
\label{P3Fourier}
\end{equation}
Differentiation of $P_3\left( u;C\right) $ with respect to $C$ yields a
relation similar to Eq.\ (\ref{DCRel}), 
\begin{equation}
\frac \partial {\partial C}P_3\left( u;C\right) =3\int_u^\infty
dx\,p_3\left( u,u,x;C\right) .  \label{P3Rel}
\end{equation}
At $C=1$, the distribution $P_3\left( u;C\right) $ reduces to the same
one-point distribution $P_1\left( u\right) $ as before, so 
\begin{equation}
P_3\left( u;C\right) =P_1\left( u\right) -3\int_C^1dC\int_u^\infty
dx\,p_3\left( u,u,x;C\right) .  \label{P3Int}
\end{equation}
The integral over $x$ in Eq.\ (\ref{P3Int}) can be evaluated by treating $%
p_3\left( u,u,x;C\right) \equiv f_1\left( x\right) $ as a one-dimensional
Gaussian distribution in $x$ (although $f_1$ is not properly normalized).
Its Fourier image $\tilde{f}_1\left( k\right) $ is easily found by
integrating Eq.\ (\ref{P3Fourier}) over $k_1$ and $k_2$: 
\begin{eqnarray}
\tilde{f}_1\left( k\right) &=&\int \int \frac{dk_1dk_2}{\left( 2\pi \right)
^2}e^{ik_1u+ik_2u}\tilde{p}_3\left( k_1,k_2,k;C\right)  \nonumber \\
&=&\frac{1}{2\pi \sqrt{1-C^2}}\exp \left( -\frac{u^2}{1+C}\right) \exp \left[
-\frac{\left( 1-C\right) \left( 2C+1\right) }{1+C}\frac{k^2}2-\frac{2uC}{1+C}%
ik\right] .  \label{P31Fourier}
\end{eqnarray}
For a general one-point Gaussian density given by its Fourier image 
\begin{equation}
\tilde{p}_1\left( k\right) =N\exp \left( -\alpha k^2/2-i\beta k\right) 
\end{equation}
one readily obtains 
\begin{equation}
\int_u^\infty dx\,p_1\left( x\right) =\frac N2\left( 1-%
\mathop{\rm erf}
\frac{u-\beta }{\sqrt{2\alpha }}\right) .  \label{P1GenF}
\end{equation}
We only need to substitute the coefficients $N$, $\alpha $ and $\beta $ from
Eq.\ (\ref{P31Fourier}) into Eqs.\ (\ref{P3Int})--(\ref{P31Fourier}) to get
an expression for $P_3$: 
\begin{equation}
P_3\left( u;C\right) =P_1\left( u\right) -3\int_C^1\frac{dx}
{4\pi \sqrt{1-x^2}}\exp \left( -\frac{u^2}{1+x}\right)
\left( 1-\mathop{\rm erf}
\frac{u\sqrt{1-x}}{\sqrt{2\left( 1+x\right) \left( 1+2x\right) }}\right) .
\label{P3Ans}
\end{equation}
As before, the integrand can be expanded in powers of $\left( 1-C\right) $
and integrated term by term.

In a similar manner it is possible to evaluate the distribution $P_3\left(
u;C_{ij}\right) $ for a general $3$-point configuration with unequal
correlations $C_{ij}$ between points. In that case, introduce a
formal dependence of $C_{ij}$ on a parameter $\epsilon $, 
\begin{equation}
C_{ij}\left( \epsilon \right) =1-\epsilon \left( 1-C_{ij}\right) , 
\end{equation}
which interpolates between the original configuration $\left( \epsilon
=1\right) $ and the degenerate case $C_{ij}=1$ at $\epsilon =0$. Then
differentiate $P_3\left( u;C_{ij}\left( \epsilon \right) \right) $
with respect to $\epsilon $ to obtain a relation analogous to Eq.\ (\ref
{P3Rel}), and subsequently integrate over $\epsilon $ from $0$ to 
$1$.

The calculation of the Euler characteristic on a hexagonal lattice as given
by Eq.\ (\ref{ChiHex}) can be further simplified because it contains
only the combination $P_1-3P_2+2P_3$. With the distribution $P_2$ given by
Eq.\ (\ref{P2Rel}), and using the identity 
\begin{equation}
p_2\left( u,u;C\right) =\int_{-\infty }^\infty dx\,p_3\left( u,u,x;C\right),
\end{equation}
one obtains 
\begin{equation}
P_1-3P_2+2P_3=3\int_C^1dC\left( \int_{-\infty }^\infty -2\int_u^\infty
\right) dx\,p_3\left( u,u,x;C\right) .
\end{equation}
In the notation of Eq.\ (\ref{P1GenF}) 
\begin{equation}
\left( \int_{-\infty }^\infty -2\int_u^\infty \right) dx\,p_1\left( x\right)
=N\mathop{\rm erf}
\frac{u-\beta }{\sqrt{2\alpha }},
\end{equation}
and we arrive at the expression 
\begin{equation}
P_1-3P_2+2P_3=3\int_C^1\frac{dx}{2\pi 
\sqrt{1-x^2}}\exp \left( -\frac{u^2}{1+x}\right) 
\mathop{\rm erf}
\frac{u\sqrt{1-x}}{\sqrt{2\left( 1+x\right) \left( 1+2x\right) }}.
\label{ChiComb}
\end{equation}
After carrying out the series expansion in $\left( 1-C\right) $ and
term-by-term integration above, we obtained the result represented in Eq.\ (%
\ref{Chi3Series}) of Sec.~\ref{Sec:Gaussian}.

Finally we turn to the calculation of the distribution $P_4\left(
u\right) $. We only treat the special case when only two of the pair
correlations between four points are independent. Namely, we assume that 
\begin{eqnarray}
\left\langle x_1x_3\right\rangle &=&\left\langle x_1x_4\right\rangle
=\left\langle x_2x_3\right\rangle =\left\langle x_2x_4\right\rangle
=B, \nonumber\\
\left\langle x_1x_2\right\rangle &=&\left\langle x_3x_4\right\rangle =C.
\end{eqnarray}
Then the Fourier image $\tilde{p}_4\left( {\bf k};B,C\right) $ of the $4$%
-point distribution density $p_4\left( {\bf x};B,C\right) $ is 
\begin{equation}
\tilde{p}_4\left( {\bf k};B,C\right) =\exp \left[ -\frac 12\left(
k_1^2+k_2^2+k_3^2+k_4^2\right) -B\left( k_1+k_2\right) \left( k_3+k_4\right)
-C\left( k_1k_2+k_3k_4\right) \right] .  \label{P4Fourier}
\end{equation}
The probability integral $P_4\left( u;B,C\right) $ is defined by Eq.\ (\ref
{PNDef}). We shall obtain an expansion of $P_4$ in $\left( 1-C\right) $
which will be useful for the case when the two pairs $\left( x_1,x_2\right) $
and $\left( x_3,x_4\right) $ of close-by points are separated by a
comparatively large distance. Differentiating with respect to $C$
gives the relation 
\begin{equation}
\frac \partial {\partial C}P_4\left( u;B,C\right) =2\int_u^\infty
dx_1\int_u^\infty dx_2\,p_4\left( x_1,x_2,u,u;B,C\right) .  \label{Rel4Diff}
\end{equation}
As before, we choose the boundary condition at $C=1$ where 
$P_4\left( u;B,1\right) =P_2\left( u;B\right)$,
and then we can integrate Eq.\ (\ref{Rel4Diff}) over $C$. The remaining task
is to evaluate the double integral in Eq.\ (\ref{Rel4Diff}), and for this we
again employ the same technique of converting two integrations into a single
one. We treat the density $p_4\left( x_1,x_2,u,u;B,C\right) \equiv f\left(
x_1,x_2\right) $ as a two-point (unnormalized) Gaussian distribution for $%
x_1 $ and $x_2$. The Fourier image of $f\left( x_1,x_2\right) $ is easily
obtained by integration of Eq.\ (\ref{P4Fourier}), 
\begin{eqnarray}
\tilde{f}\left( k_1,k_2\right) &=&\int \frac{dk_3}{2\pi }e^{iuk_3}\int \frac{%
dk_4}{2\pi }e^{iuk_4}\,\tilde{p}_4\left( k_1,k_2,k_3,k_4;B,C\right) 
\nonumber \\
&=&N\exp \left[ -\frac \alpha 2\left( k_1^2+k_2^2\right) -\beta
k_1k_2-i\gamma \left( k_1+k_2\right) \right] ,
\end{eqnarray}
where 
\begin{equation}
N\equiv \frac{1}{2\pi \sqrt{1-B^2}}\exp \left( -{u^2\over 1+B}\right), 
\qquad\alpha \equiv 1-\frac{2B^2}{1+C},
\qquad\beta \equiv C-\frac{2B^2}{1+C},
\qquad\gamma \equiv \frac{2Bu}{1+C}.
\end{equation}
The two-point density $N^{-1}f\left( x_1,x_2;\alpha ,\beta ,\gamma \right) $
describes a Gaussian distribution with nonzero mean $\left\langle
x_i\right\rangle =\gamma $, variance $\left\langle x_i^2\right\rangle
=\alpha $, and correlation $\left\langle x_1x_2\right\rangle =\beta $, and
is similar to $p_2\left( x,y;C\right) $. Then we can directly use Eqs.\ (\ref
{P1MinusP2})--(\ref{IntSer2}) in which we must replace $C$ by $\beta /\alpha 
$ and $u$ by $\left( u-\gamma \right) /\sqrt{\alpha }$ and also multiply by $%
N$: 
\begin{equation}
\int_u^\infty dx\int_u^\infty dy\,f\left( x,y;\alpha ,\beta ,\gamma \right) =%
\frac N2\left( 1-%
\mathop{\rm erf}
\frac{u-\gamma }{\sqrt{2\alpha }}\right) -\int_{\beta /\alpha }^1\frac{Ndx}{%
2\pi \sqrt{1-x^2}}\exp \left( -\frac{\left( u-\gamma \right) ^2}{\alpha
\left( 1+x\right) }\right) .  \label{GenIntf2}
\end{equation}
The resulting series in $\epsilon \equiv \left( 1-\beta /\alpha \right) $
can then be converted into a series in $\left( 1-C\right) ,$ substituted
into Eq.\ (\ref{Rel4Diff}) and integrated term by term over $C$. Because the
resulting expressions are quite cumbersome, we do not write them out 
explicitly.

Similarly, one can obtain an expansion of $P_4\left( u\right) $ for the case
when all four points are close to each other and the parameters $B$ and $C$
approach $1$ simultaneously; such an expansion is necessary for the
calculation of the Euler characteristic on a square lattice. By
parametrizing $B=1-\delta $ and $C=1-q\delta $, where $q$ is a formal
constant parameter with value of order $1$, and differentiating $P_4$ with
respect to $\delta $, one obtains a relation similar to Eq.\ (\ref{Rel4Diff}%
), 
\begin{equation}
\frac \partial {\partial \delta }P_4\left( u;B,C\right) =-\int_u^\infty
dx_1\int_u^\infty dx_2\,\left[ 2qp_4\left( x_1,x_2,u,u;B,C\right)
+4p_4\left( x_1,u,x_2,u;B,C\right) \right] .
\end{equation}
The double integrals are treated in the same manner as above, expanded in $%
\delta $ and then integrated term by term. The actual values of $B$ and $C$
can be then substituted into the resulting series in $\delta $. In this
case, however, the series (\ref{IntSer2}) cannot be used directly because in
the limit of $\delta =0$ the parameters $\alpha $, $\beta $, $\gamma $ all
tend to zero while the expansion parameter $\epsilon \equiv \left( 1-\beta
/\alpha \right) $ of Eq.\ (\ref{IntSer2}) does not become small. Instead,
the parameters $\alpha $, $\beta $, $\gamma $ can be expressed in
terms of
$\delta $ and then a direct expansion of Eq.\ (\ref{GenIntf2}) 
in $\delta $ can be performed; the
integrations for each term of that expansion can be calculated analytically.
Again, we do write out the explicit form of the distribution $P_4$.

\end{document}